\documentclass[prd,aps,preprint,onecolumn,superscriptaddress,tightenlines,nofootinbib,eqsecnum,preprintnumbers,longbibliography,12pt]{revtex4-1}
\pdfoutput=1
\usepackage{amsmath,latexsym,amssymb,graphicx,ifpdf,slashed,color,hyperref,url,cancel}
\usepackage[T1]{fontenc}
\usepackage{relsize}
\hypersetup{colorlinks,citecolor= red ,linkcolor= niceblue, urlcolor=niceblue}
\definecolor{niceblue}{rgb}{0.1,0.2,0.6}

\begin{document}
\def\Berkeley{Department of Physics, University of California Berkeley, Berkeley, CA 94720, USA}
\def\Carleton{Department of Physics, Carleton University, Ottawa, ON K1S 5B6, Canada}
\def\Fermilab{Theoretical Physics Department, Fermilab, P.O. Box 500, Batavia, IL 60510, USA}
\def\Loyola{Department of Physics, Loyola University Chicago, Chicago, IL 60660, USA}
\def\Northwestern{Department of Physics and Astronomy, Northwestern University, Evanston, IL 60208, USA}

\title{
{Origin of Sterile Neutrino Dark Matter via\\ Secret Neutrino Interactions with Vectors}}
\author{Kevin J. Kelly}
\email{kkelly12@fnal.gov}
\affiliation{\Fermilab}

\author{Manibrata Sen}
\email{manibrata@berkeley.edu }
\affiliation{\Berkeley}
\affiliation{\Northwestern}

\author{Walter Tangarife}
\email{wtangarife@luc.edu}
\affiliation{\Loyola}

\author{Yue Zhang}
\email{yzhang@physics.carleton.ca}
\affiliation{\Carleton}

\date{\today}

\begin{abstract}
Secret neutrino interactions can play an essential role in the origin of dark matter. We present an anatomy of production mechanisms for sterile neutrino dark matter, a keV-scale gauge-singlet fermion that mixes with active neutrinos, in the presence of a new vector boson mediating secret interactions among active neutrinos. We identify three regimes of the vector boson's mass and coupling where it makes distinct impact on dark matter production through the dispersion relations and/or scattering rates. We also analyze models with gauged $L_\mu-L_\tau$ and $B-L$ numbers which have a similar dark matter cosmology but different vector boson phenomenology. We derive the parameter space in these models where the observed relic abundance is produced for sterile neutrino dark matter. They serve as well-motivated target for the upcoming experimental searches.
\end{abstract}

\preprint{FERMILAB-PUB-20-181-T, NUHEP-TH/20-02}

\maketitle
\tableofcontents

\section{Introduction}
The nature of dark matter is one of the biggest puzzles of our universe. Many experimental searches for dark matter are ongoing but with no success yet. New theoretical targets are needed to guide the path forward. Neutrinos, in many aspects, are the least well-known particles within the Standard Model owing to their weak interacting nature. In particular, neutrino self-interactions have never been directly measured in laboratories. Only indirect bounds have been inferred~\cite{Bilenky:1992xn,Bilenky:1994ma,Masso:1994ww,Bilenky:1999dn,Belotsky2001,Ng:2014pca,Ioka:2014kca,Das:2017iuj,Berryman:2018ogk,Kelly:2018tyg,Kelly:2019wow,Kreisch:2019yzn,Blinov:2019gcj,deGouvea:2019qaz,Shalgar:2019rqe,Bustamante:2020mep,Brdar:2020nbj,Deppisch:2020sqh}. There is room for neutrinos to interact with themselves much more strongly than the ordinary weak interaction. This also accommodates a potential connection to dark matter. We investigate such a possibility in the present work.

Sterile neutrino, a gauge singlet fermion that mixes with the Standard Model neutrinos, is one of the simplest dark matter candidates. It is requires very little effort of model building to realize such a candidate. Its cosmological longevity does not even require a symmetry but is simply attributed to the small mass and mixing. As fermionic dark matter, a sterile neutrino cannot be arbitrarily light~\cite{Tremaine:1979we,Boyarsky:2008ju}, which grants the opportunity to observe monochromatic photons from its decay as an indirect detection signal~\cite{Pal:1981rm,Abazajian:2001nj}.
The attractiveness of sterile neutrino dark matter is further enhanced by a novel finding of how its relic abundance could be produced.
Dodelson and Widrow~\cite{Dodelson:1993je} pointed out that the same mixing angle allowing sterile neutrino to decay also enables it to be efficiently produced in the early universe.
The corresponding parameter space has been scrutinized by various astrophysical probes and is virtually ruled out by the existing constraints~\cite{Watson:2011dw, Horiuchi:2013noa, Perez:2016tcq, Dessert:2018qih, Ng:2019gch}. A minimal solution to ameliorate this tension is to allow for a non-zero lepton-number asymmetry~\cite{Shi:1998km}. However, the required asymmetry does not have to be large and is challenging to observe elsewhere~\cite{Abazajian:2004aj}. Alternative proposals include inflaton decays~\cite{Shaposhnikov:2006xi}, freeze-in via additional particles~\cite{Asaka:2006ek, Roland:2014vba}, new interactions in the sterile sector~\cite{Hansen:2017rxr, Johns:2019cwc}, and gauge extensions of the Standard Model~\cite{Bezrukov:2009th, Nemevsek:2012cd, Dror:2020jzy}.
These mechanisms often make dramatic changes to the ultraviolet side of the story, i.e., the initial conditions of the early universe.

The focus of this work are theories that keep the most salient feature of the Dodelson-Widrow mechanism, the infrared dominance of dark matter production, 
and make testable predictions experimentally. The intuition is quite simple. Dark matter relic abundance is set by the product of the two ingredients in this mechanism, active-sterile neutrino mixing and the active neutrino reaction rate. 
While the former is strongly constrained by indirect searches, the latter is allowed to be much stronger than the weak interaction, especially when neutrinos are interacting with themselves. This idea was recently entertained within a model where neutrinos self-interact via a lepton-number charged scalar particle~\cite{deGouvea:2019phk}.
In the present work, we explore the impact of a new vector boson with coupling to neutrinos on the production of sterile neutrino dark matter.
During the dark matter production epoch, such a vector boson could manifest as a heavy mediator, or a light and thermalized degree of freedom.
These possibilities help us to discover new variations beyond the original Dodelson-Widrow mechanism.
We consider three incarnations of the vector boson, in a neutrinophilic model, as well as in models with gauged ${L_\mu-L_\tau}$ and ${B-L}$ symmetries.
We carry out detailed calculations of the relic density of sterile neutrino dark matter in each model, and derive the parameter space for the observed value to be reproduced. Furthermore, we confront our results with the existing and future experimental searches for each type of the new vector boson.

This paper is organized as follows. In the next section, we give an overview of sterile neutrino dark matter, and use the tension between Dodelson-Widrow and astrophysical constraints as the motivation for introducing new neutrino forces. In section~\ref{sec:neutrinphilic}, we first analyze a model with a neutrinophilic vector boson, deriving the parameter space for dark matter relic abundance and comparing it with a number of experimental probes. In sections~\ref{sec:LmuLtau} and \ref{sec:B-L}, we do the same but in the gauged $U(1)_{L_\mu-L_\tau}$ and $U(1)_{B-L}$ models, respectively.
We draw conclusions in section~\ref{sec:conclude}.

\section{Motivation for Secret Neutrino Interactions}\label{sec:mechanism}
We start with the following Lagrangian which introduces a sterile neutrino field,
\begin{equation}\label{eq:Sterile}
\mathcal{L}_{{\rm SM}+\nu_s} = \mathcal{L}_{\rm SM} + \bar \nu_s i \cancel\partial  \nu_s - 
\left[ y_\varepsilon \nu_s^T C^{-1} H^T (i\sigma_2) L  + {\rm h.c.} \rule{0mm}{4mm}\right]  \ ,
\end{equation}
where $\nu_s$ is a left-handed fermion and a Standard Model gauge singlet, $C= -i\gamma^2\gamma^0$, and
$L^T = (\nu, \ell^-)$ is the Standard Model lepton doublet with hypercharge $-1/2$. The parameter $y_\varepsilon$ is a tiny Yukawa coupling between $\nu_s$ and $L$ that makes a negligible contribution to the observed neutrino mass.  The main role of this term is to generate a mixing between the active and sterile neutrinos, after electroweak symmetry breaking, $\langle H\rangle^T= (0, v/\sqrt2)$.

In the case where both active and sterile neutrinos are Majorana particles, their mass terms take the form,
\begin{equation}
\frac{1}{2} \begin{pmatrix}
\nu^T & \nu_s^T
\end{pmatrix} C^{-1}
\begin{pmatrix}
m_\nu & m_d \\
m_d & m_s
\end{pmatrix}
\begin{pmatrix}
\nu \\ 
\nu_s
\end{pmatrix} + {\rm h.c.} \ ,
\end{equation}
where $m_d = y_\varepsilon v/\sqrt2$ is generated by the Yukawa coupling in Eq.~(\ref{eq:Sterile}), and $m_\nu$ and $m_s$ are the Majorana masses generated from lepton-number-violating sources beyond the above Lagrangian. Assuming the mass hierarchy $m_d^2/m_s \ll m_\nu \ll m_s$, the active-sterile neutrino mixing angle is approximately $\theta \simeq m_d/m_s$. The two mass eigenvalues are $m_1\simeq m_\nu$ and $m_4\simeq m_s$, respectively.
The heavier mass eigenstate and dark matter candidate is the following linear combination,
\begin{equation}
\nu_4 = \nu_s \cos\theta + \nu \sin\theta \ .
\end{equation}

Alternatively, if the neutrino masses are Dirac (both active and sterile), we must introduce partner fields for $\nu$ and $\nu_s$ for writing down the corresponding Dirac mass terms,
\begin{equation}
\begin{pmatrix}
N^T & \nu_s^T
\end{pmatrix}
C^{-1} \begin{pmatrix}
m_\nu & 0 \\
m_d & m_s
\end{pmatrix} 
\begin{pmatrix}
\nu \\ \nu_s'
\end{pmatrix}
+ {\rm h.c.} \ ,
\end{equation}
where $m_d = y_\varepsilon v/\sqrt2$, and $m_\nu$ and $m_s$ are Dirac masses. We set the (12) element to zero for simplicity.
In this case, $\nu$ mixes with the $\nu_s'$ field, the resulting dark matter state is $\nu_4=\nu_s' \cos\theta + \nu \sin\theta$, where $\theta \simeq m_d/m_s$. Its mass is approximately $m_4 \simeq m_s$, assuming that $m_d\ll m_s$.

In both cases, when $\theta \ll1$, the mass eigenstate $\nu_4$ has a large sterile neutrino component (i.e., $\nu_4 \simeq \nu_s^{(\prime)}$) and a very small active component.
Hereafter, we refer to $\nu_4$ as the sterile neutrino. 
The orthogonal linear combination, denoted by $\nu_1$, is a mostly-active neutrino mass eigenstate.

With a mass between keV to MeV scales, $\nu_4$ is a viable dark matter candidate. 
The angle $\theta$, parametrizing the presence of its active component, is assumed to be small enough so that $\nu_4$ never reaches thermal equilibrium with the Standard Model sector in the early universe. Assuming the universe was born without a population of $\nu_4$, its relic density could be produced through neutrino oscillation effects.
In the early universe, at temperatures before the decoupling of weak interaction, active neutrinos are constantly produced and destroyed by Standard Model weak interactions, as the flavor eigenstate $\nu$. 
Since the produced $\nu$ is a linear combination of mass eigenstates $\nu_{1,4}$, neutrino oscillations will occur.
In the early universe, the active neutrino encounters a refractive potential due to interaction with the thermal bath, which alters its dispersion relation from the zero temperature case, thereby changing the oscillation probabilities. 
However, as the neutrino oscillates on a timescale given by its energy and the mass-squared difference, it can undergo weak interactions, that reset only the active component.
By this time, the original $\nu$ state has already developed a $\nu_s$ component, which survives such a ``measurement'', and eventually contributes to the relic density of $\nu_4$ dark matter.
The above oscillation process can repeat for many times until weak interaction decouples.
This is the crux of the Dodelson-Widrow mechanism~\cite{Dodelson:1993je}.
It was demonstrated that a proper choice of the mixing parameter $\theta$ can produce the correct dark matter relic density.

\begin{figure}[t]
\begin{center}
\includegraphics[scale=0.6]{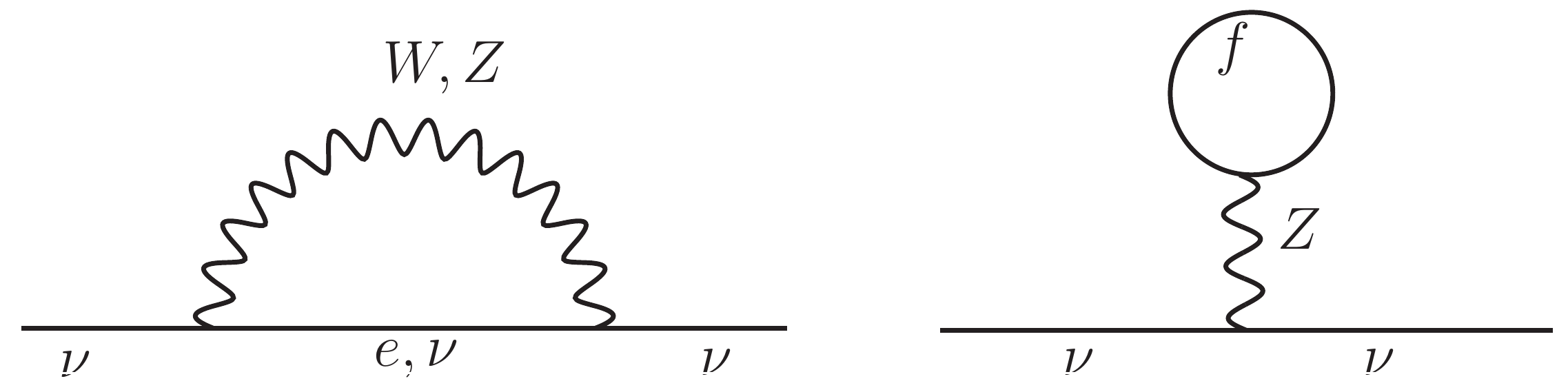}
\end{center}
\caption{Self-energy diagrams contributing to $V_T$ in the Standard Model. In the absence of a lepton asymmetry, only the left diagram contributes.}
\label{fig:neutrinophilicZW}
\end{figure}

The probability for each active neutrino $\nu$ to oscillate into a $\nu_s$ (the latter mostly ends up as dark matter $\nu_4$) is dictated by the effective mixing angle
\begin{equation}
\label{mixinganglematt}
\sin^22\theta_{\rm eff}  \simeq   \frac{\Delta^2 \sin^22\theta}{\Delta^2 \sin^22\theta + (\Gamma/2)^2 + (\Delta \cos2\theta - V_T)^2} \ ,
\end{equation}
which is obtained by taking the thermal average of time-dependent $\nu\to\nu_s$ oscillation probability.
Here, $\Delta \equiv (m_4^2 -m_1^2)/(2E) \simeq m_4^2/(2E)$ is the vacuum neutrino oscillation frequency, and $E$ is the energy of the oscillating neutrino state.
If the active neutrinos participate only in the Standard Model weak interaction, as assumed in the original work by Dodelson and Widrow, the effective thermal potential $V_T$ results from the self-energy of the neutrino, as depicted by the diagrams in Fig.~\ref{fig:neutrinophilicZW}. The resultant potential is given by \cite{Abazajian:2001nj,Abazajian:2005gj}
\begin{equation}\label{Eq:SM_VT}
V_{T,\, {\rm SM}} \,= \,- 3.72\,G_F E\,T^4 \left(\frac{2}{M_W^2}+ \frac{1}{M_Z^2} \right),
\end{equation}
where $G_F$ is Fermi's constant and $T$ is the temperature of the universe. 
The label ``SM'' indicates that this expression is only valid in the absence of any new neutrino interaction beyond the Standard Model (BSM). This is the result in the absence of any lepton number asymmetry.

The interaction or collision rate $\Gamma$, due to exchange of $W$ and $Z$ bosons, is given by \cite{Abazajian:2001nj,Abazajian:2005gj}
 \begin{equation}\label{Eq:SM_rate}
 \Gamma_{\rm SM} \simeq \left \lbrace \begin{array}{ll}
 1.27\, G_F^2 E\,T^4 & \quad {\rm for} \,\,\nu_e ,\\
 0.92\, G_F^2 E\,T^4 & \quad {\rm for} \,\,\nu_\mu,\, \nu_\tau.
 \end{array} \right.
 \end{equation}  
We will derive new expressions of these quantities in BSM frameworks in the upcoming sections.

Assuming the initial dark matter abundance to be negligible, the amount dark matter produced is given by the integration of active neutrino production rate times the oscillation probability over the cosmological time.
Because the oscillation probability is neutrino energy dependent, we write down the equation that governs the phase-space density evolution for the sterile neutrino with fixed energy $E$~\cite{Dodelson:1993je, Abazajian:2005gj, Hansen:2017rxr}
\begin{equation}
\label{masterequation}
\frac{d\, f_{\nu_4}(x, z)}{d \ln z} = \frac{\Gamma }{4H} \, \sin^22\theta_{\rm eff}\, f_{\nu}(x) \,,
\end{equation}
where $H$ is the Hubble parameter and the parameter $z \equiv \mu/T$ is introduced to label the cosmological time. Clearly, the choice of $\mu$ does not affect the result, and for convenience, we choose  $\mu \equiv 1\,$MeV throughout this work.
$f_{\nu}$ is the Fermi-Dirac distribution for an active neutrino or antineutrino, of the form $f_{\nu}(x) = 1/(1+e^x)$, where $x\equiv E/T$ and it is $z$-independent.
Unlike the proposal of~\cite{Shi:1998km}, we will not consider the presence of a lepton asymmetry in the universe, thus the chemical potential for active neutrino is set to zero.

This is an elegant mechanism, given its simplicity and predictability. Indeed, sterile neutrino dark matter has been explored extensively, from the model building to its various aspects in detection (see e.g., \cite{Boyarsky:2009ix, Abazajian:2017tcc}), and strong tensions emerge from the indirect detection. In particular, a nonzero $\nu-\nu_s$ mixing $\theta$ makes the sterile neutrino dark matter $\nu_4$ unstable. It could decay into either $\nu_1$ plus a photon, or three $\nu_1$. The former decay channel could lead to extra $X$-ray radiation from regions of accumulating dark matter. The non-observation by $X$-ray telescopes disfavors the mixing angle $\theta$ needed for the Dodelson-Widrow mechanism to work in regions where the sterile neutrino  is heavier than a few keV.
Meanwhile, as a fermionic dark matter, it is found difficult to successfully fill lighter sterile neutrino into the known dwarf galaxies~\cite{Tremaine:1979we, Boyarsky:2008ju, Gorbunov:2008ka}. As a result, the entire parameter space that produces the correct relic density for sterile neutrino dark matter via the Dodelson-Widrow mechanism is almost closed~\cite{Watson:2011dw, Horiuchi:2013noa, Perez:2016tcq, Dessert:2018qih, Ng:2019gch}.

It is also worth noting that there have been claims of an unresolved $\sim 3.55\,{\rm keV}$ X-ray line from various nearby galaxy clusters~\cite{Bulbul:2014sua, Boyarsky:2014jta}, which is under thorough scrutinization nowadays~\cite{Dessert:2018qih, brawl,Abazajian:2020unr,Boyarsky:2020hqb}. While the dust has not settled, the relevant takeaway for our work is that if sterile neutrino dark matter decays into this line, the corresponding mixing angle $\theta$ is too small to account for its relic density via the Dodelson-Widrow mechanism.

A recent letter~\cite{deGouvea:2019phk} has suggested that a new, secret interaction among the active neutrinos is effective for alleviating the above tensions. 
There, it was postulated that the secret neutrino interaction is mediated by a lepton-number charged scalar and much stronger than the ordinary weak interaction, thus allowing the sterile neutrinos to be produced more efficiently in the early universe. It was shown that with the new interactions, one can explain the relic density without violating all the existing constraints,
for dark matter mass between a few keV to MeV scale.
This includes the point favored by the observation of the $3.55\,{\rm keV}$ X-ray line.
As the most intriguing aspect of this model, it addresses the relic density of dark matter but with the new physics being introduced in the active neutrino sector.
There are predictions in low energy experiments, such as precision measurement of pion, kaon decays, and accelerator neutrino facilities with a near detector (e.g., DUNE). In turn, these probes of secret neutrino interaction can test the fate of dark matter.

In this work, we take this enticing idea further by analyzing models for secret neutrino interactions mediated by mediated by a new vector boson $V$.
Such a new vector boson could naturally arise from $U(1)$ gauge extensions of the Standard Model. 
In the upcoming sections, we explore three of its incarnations: {\bf i)} \hyperref[sec:neutrinphilic]{a model with a neutrinophilic $V$}; {\bf ii)} \hyperref[sec:LmuLtau]{gauged $U(1)_{L_\mu-L_\tau}$ model}; and {\bf iii)} \hyperref[sec:B-L]{gauged $U(1)_{B-L}$ model}. In each case, we confront the parameter space favored by relic density with existing and future experimental constraints.

Strictly speaking, in the last two models, the neutrino interactions are not so secret because other Standard Model particles also see them, and naively they are already tightly constrained. 
Interestingly, we find that there still exists a viable parameter space to accommodate the correct dark matter relic density  in these models. Future experiments will fully probe the remaining parameter space, to either discover or falsify our proposal.
In contrast, the neutrinophilic vector boson model with a genuine secret neutrino self interaction is less constrained and calls for new experiments for it to be fully covered.

\section{Model with a Neutrinophilic Vector Boson} \label{sec:neutrinphilic}
In the first model, we consider a new vector boson $V$ which couples only to the active neutrinos. Because neutrinos exist in an $SU(2)_L$ doublet, such a neutrinophilic nature of $V$ is achieved via a higher dimensional operator, added to the Lagrangian in Eq.~(\ref{eq:Sterile}),
\begin{equation}\label{eq:Phen}
\mathcal{L} =  \mathcal{L}_{{\rm SM}+\nu_s} - \frac{1}{4} V_{\mu\nu} V^{\mu\nu} + \frac{1}{2} m_V^2 V_\mu V^\mu + \sum_{\alpha, \beta=e,\mu,\tau} \frac{(\overline{L}_\alpha i\sigma_2 H^*) \gamma_\mu (H^T i\sigma_2 L_\beta) V^\mu}{\Lambda^2_{\alpha\beta}} \ ,
\end{equation}
where the cutoff scale $\Lambda_{\alpha\beta}$ characterizes the interaction strength of $V$ with different combinations of lepton flavors. After electroweak symmetry breaking, the vacuum expectation value of the Higgs boson projects out the neutrino field from the lepton doublets. At low energies, the $V$ couplings are neutrinophilic. 
\begin{equation}
\mathcal{L}_{\nu\bar\nu V} = \sum_{\alpha, \beta=e,\mu,\tau} \lambda_{\alpha\beta} \bar\nu_\alpha \gamma^\mu \nu_\beta V_\mu \ ,
\end{equation}
where the couplings are
$\lambda_{\alpha\beta} = v^2/(2\Lambda^2_{\alpha\beta})$.

Such a new interaction, if strong enough, could keep the neutrinos in thermal equilibrium with themselves longer than the weak interaction, thus facilitating the production rate of sterile neutrino dark matter in the early universe.
The task of this section is to quantify this statement and find the favored model parameter space for which the sterile neutrino has a relic density that matches today's observed amount.
Meanwhile, the neutrinophilic vector boson could also lead to observational effects in various processes in the laboratories where neutrinos interact.
In the  following subsections, we derive a list of existing and near-future experimental coverage on the model parameter space which have interesting interplay with the relic density favored region. 
Up to subsection~\ref{sec:DUNE}, we consider a representative case where $V$ couplings only to the muon neutrino $\nu_\mu$, 
and the sterile neutrino also mixes with $\nu_\mu$. 
The flavor dependence of our analysis will be commented afterwards, where it is pointed out that the relic density results remain similar for other choices of flavors.
Subsection~\ref{subsec:DifferentFlavors} comments on the phenomenological implications of allowing $V$ to couple to different flavors of neutrinos.
In subsection~\ref{subsec:NeutrinophilicUV}, we present a possible ultraviolet (UV) completion for the effective operator introduced in Eq.~(\ref{eq:Phen}).

\subsection{The anatomy of sterile neutrino dark matter production}\label{sec:anatomy}
In the original Dodelson-Widrow mechanism, the relic density of sterile neutrino dark matter depends on the neutrino weak interaction rate $\Gamma$, and the effective active-sterile neutrino mixing angle that is also controlled by the weak interaction, through the thermal potential $V_T$. In the presence of the new neutrino self interaction mediated by $V$, both $V_T$ and $\Gamma$ are modified, although the relic density can still be calculated using Eq.~(\ref{masterequation}).

\begin{figure}[!t]
\begin{center}
\includegraphics[scale=0.6]{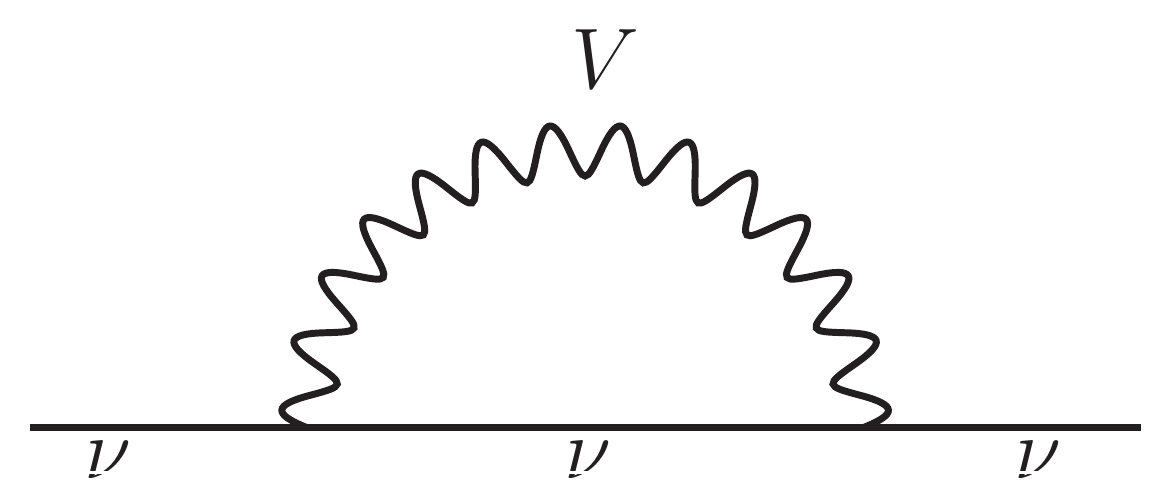}\quad
\end{center}
\caption{Loop contribution of the new interaction to the active-neutrino thermal potential $V_T$.}
\label{fig:neutrinophilicV}
\end{figure}
The new contribution to the active-neutrino thermal potential is shown in Fig.~\ref{fig:neutrinophilicV}, and takes the following form~\cite{Notzold:1987ik, Quimbay:1995jn}:
\begin{equation}\label{eq:VT}
{\small \begin{split}
V_{T,\, V}(E, T) &= \frac{|\lambda_{\mu\mu}|^2}{8\pi^2 E^2} \int_{0}^\infty dp \left[ \left( \frac{m_V^2 p}{2\omega} L_2^+(E, p) 
- \frac{4 E p^2}{\omega} \right)\frac{1}{e^{\omega/T}-1} + \left( \frac{m_V^2}{2} L_1^+(E, p) - 4 E p \right) \frac{1}{e^{p/T}+1} \right] \ , \\
L_1^+ (E, p) & = \ln \frac{4 p E + m_V^2}{4 p E - m_V^2}, \ \ \ \ \ 
L_2^+ (E, p)  = \ln \frac{\left( 2 p E + 2 E \omega + m_V^2 \right)\left(2 p E - 2 E \omega + m_V^2 \right)}
{\left( -2 p E + 2 E \omega + m_V^2 \right) \left( -2 p E - 2 E \omega + m_V^2 \right)} \ ,
\end{split}}
\end{equation}
where $\omega=\sqrt{p^2+m_V^2}$.
In the very heavy or very light mediator $V$ limit, the above potential simplifies into
\begin{equation}\label{eq:VT2}
V_{T,\, V}(E, T) = \left\{ \begin{array}{ll}
- 7 \pi^2 |\lambda_{\mu\mu}|^2 E T^4/(45 m_V^4), & \hspace{1cm} T \ll m_V \\
+ |\lambda_{\mu\mu}|^2 T^2/(8 E), & \hspace{1cm} T \gg m_V\, .
\end{array} \right.
\end{equation}
It is worth noting that this potential changes sign as $T$ passes $m_V$.
These asymptotic expressions are useful for us to infer the parametric dependence in the final relic density.
In the numerical integration of the Boltzmann equation, we keep the most general form of the thermal potential, Eq.~(\ref{eq:VT}).

\begin{figure}[t]
\begin{center}
\includegraphics[scale=0.5]{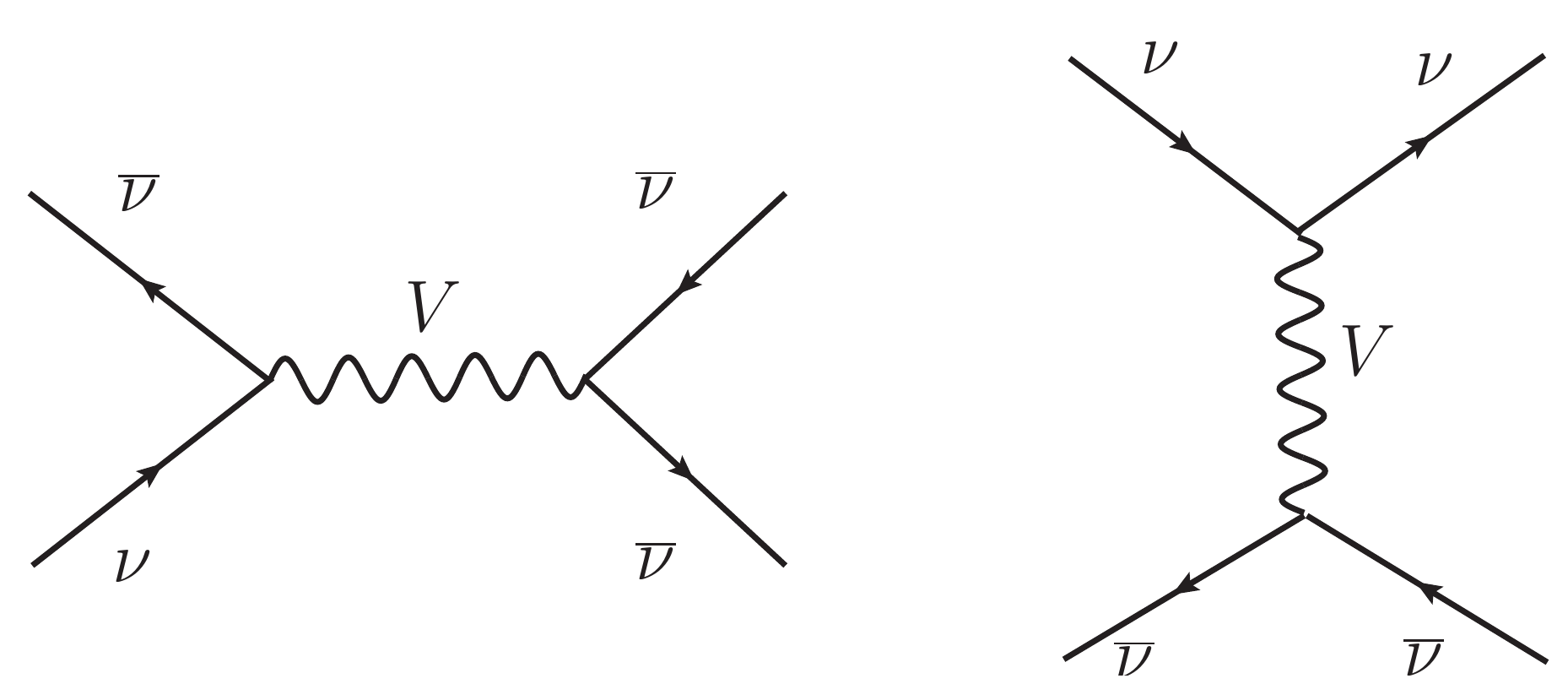}\\
\vspace{0.1in}
\includegraphics[scale=0.5]{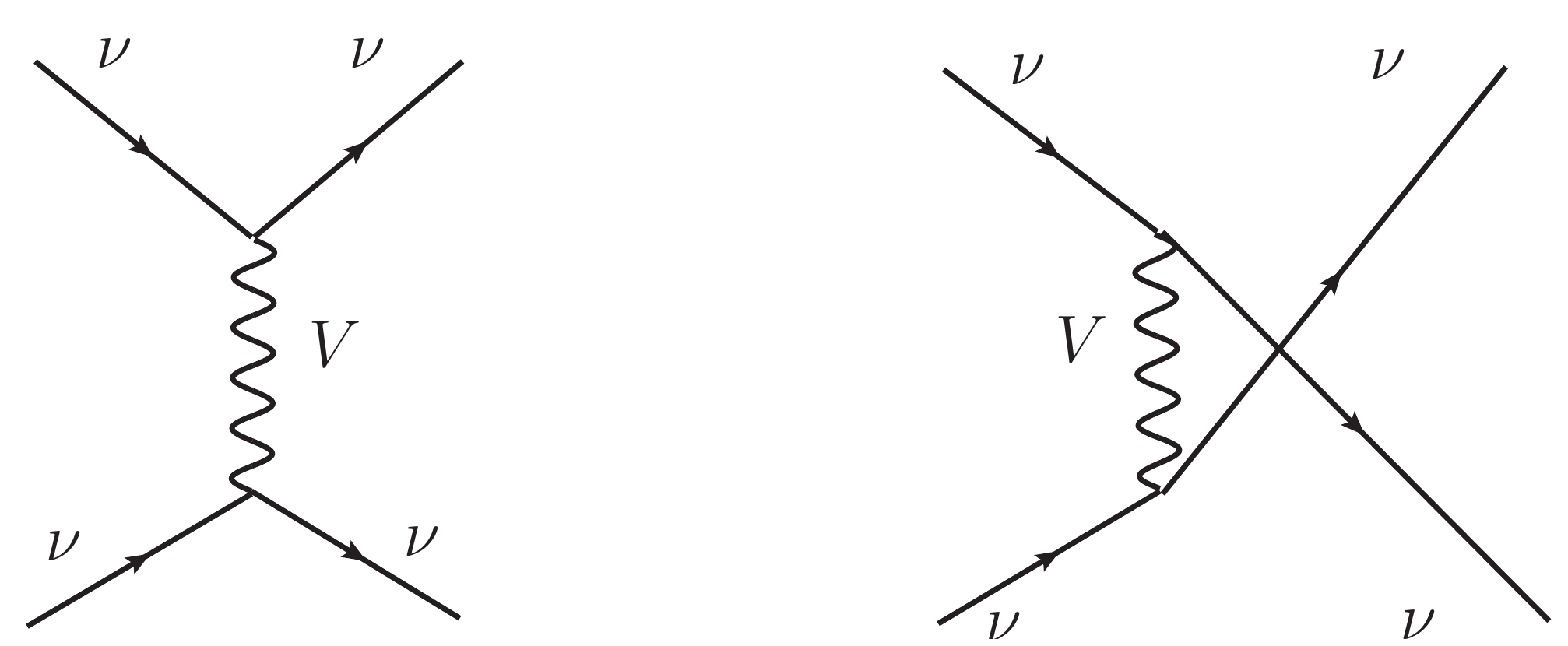}
\end{center}
\caption{Scattering diagrams contributing to $\Gamma_V$ . The top panel represents contribution to the $\nu \bar \nu \to \nu \bar \nu$ channel, whereas the bottom panel shows the contribution to the $\nu \nu \to \nu  \nu$.  For $m_V\lesssim T$, the leading diagram is that of $\nu_\mu \bar \nu_\mu \to \nu_\mu \bar \nu_\mu$ with an on-shell $V$ exchange.  }
\label{fig:Scattering}
\end{figure}

The new interaction mediated by the $V$ boson provides new scattering channels among the neutrinos, including
$\nu_\mu \nu_\mu \to \nu_\mu \nu_\mu$ and $\nu_\mu \bar \nu_\mu \to \nu_\mu \bar \nu_\mu$, as shown by the Feynman diagrams in Fig.~\ref{fig:Scattering}.
The corresponding cross sections are
\begin{equation}\label{eq:cross}
\begin{split}
\sigma_{\nu_\mu \nu_\mu \to \nu_\mu \nu_\mu} &= \frac{|\lambda_{\mu\mu}|^4}{4\pi m_V^2} 
\left[ \frac{s}{s+m_V^2} + \frac{2 m_V^2}{s + 2 m_V^2} \log\left( 1 + \frac{s}{m_V^2} \right) \right] \ , \\
\sigma_{\nu_\mu \bar \nu_\mu \to \nu_\mu \bar \nu_\mu} &= \frac{|\lambda_{\mu\mu}|^4}{4\pi (s - m_V^2)^2}
\left[ \frac{s^2}{m_V^2} - 4 m_V^2 + \frac{4 \left( m_V^4 - s^2 \right)}{s}  \log\left( 1 + \frac{s}{m_V^2} \right)
+ \frac{10 s}{3} \right] \ ,
\end{split}
\end{equation}
where $\sqrt{s}$ is the center of mass energy.
If the initial-state four-momenta are denoted as $(E, \vec{p})$ and $(E', \vec{p}')$, we have $s=2 E E' - 2 \vec{p}\cdot \vec{p}' = 2 E E' (1-\cos\vartheta)$, where the tiny active neutrino masses are neglected, and $\vartheta$ is the relative angle between $\vec{p}$ and $\vec{p}'$.

These cross sections are used for calculating the thermal averaged reaction rates for a neutrino with fixed energy $E$, by averaging over the 
phase space of the other neutrino (or anti-neutrino) it scatters with,
\begin{equation} \label{Eq:Gamma_def}
\Gamma_V (E, T) = 2 \int \frac{d^3\vec{p}'}{(2\pi)^3}\ f_{\nu}(E',T)\ \sigma_{\rm tot}(\vec{p}, \vec{p}')\ v_\text{Møller} \ , 
\end{equation}
where $f_{\nu}(E',T) = 1/[1+\exp(E'/T)]$ is the Fermi-Dirac distribution for active neutrino or antineutrino, 
$\sigma_{\rm tot}$ is the sum of the two cross sections in Eq.~(\ref{eq:cross}),
and the Møller velocity is equal to $(1-\cos\vartheta)$ for ultra-relativistic neutrinos.

Here, $\Gamma_V$ also includes the reaction rate of antineutrino, which occurs via the charge-conjugate channels of those reported in Eq.~(\ref{eq:cross}).
At high temperatures when helicity is approximately equivalent to chirality, neutrinos are left-handed and antineutrinos are right-handed. 
Both of them can contribute to the dark matter relic density via the mixing and oscillation (in a helicity preserving way).
Thus, the prefactor of 2 captures both helicity states of the active neutrinos and antineutrinos.

It is useful to show the asymptotic forms of the cross sections and thermal rates in the heavy or light $V$ limits.
For $m_V\gg T$, we find
\begin{equation}
\sigma_{\nu_\mu \bar \nu_\mu \to \nu_\mu \bar \nu_\mu} =  \frac{|\lambda_{\mu\mu}|^4 \,s}{3\pi m_V^4}, \qquad 
\sigma_{\nu_\mu \nu_\mu \to \nu_\mu \nu_\mu} =  \frac{|\lambda_{\mu\mu}|^4\, s}{2\pi m_V^4} \ ,
\end{equation} 
and
\begin{equation}
\Gamma_V^\text{low-T}(E, T) = \frac{7\pi |\lambda_{\mu\mu}|^4 E T^4}{54 m_V^4} \ .
\end{equation} 
In contrast, for $T\gg m_V$, the typical center-of-mass energy of the scattering is much higher than $m_V$. In this limit,
the cross sections in Eq.~(\ref{eq:cross}) takes the forms
\begin{equation}
\sigma_{\nu_\mu \bar \nu_\mu \to \nu_\mu \bar \nu_\mu} = \sigma_{\nu_\mu \nu_\mu \to \nu_\mu \nu_\mu} =  \frac{|\lambda_{\mu\mu}|^4}{4\pi m_V^2} \ .
\end{equation} 
These cross section blow up in the $m_V\to0$ limit, similar to the case of Bhabha scattering in QED. Here the scattering range is regularized by a non-zero $V$ mass.
Adding the two, the corresponding thermal rate is
\begin{equation}
\label{highTrate}
\Gamma_V^\text{high-T}(E, T) = \frac{3 \zeta(3) |\lambda_{\mu\mu}|^4 T^3}{4\pi^3 m_V^4} \ .
\end{equation} 
In addition, for $T\gtrsim m_V$, the process $\nu_\mu \bar \nu_\mu \to \nu_\mu \bar \nu_\mu$ could also occur with an on-shell $V$ exchange. 
In the early universe, this can also be interpreted as the decay and inverse decay of fully thermalized $V$ particles. In the narrow width approximation,
\begin{equation}
\sigma_{\nu_\mu \bar \nu_\mu \to \nu_\mu \bar \nu_\mu} \simeq \frac{|\lambda_{\mu\mu}|^4 m_V}{12 \gamma_V} \delta(s-m_V^2) \ ,
\end{equation}
where $\gamma_V$ is the  decay rate of $V$, $\gamma_V = |\lambda_{\mu\mu}|^2 m_V /(12\pi)$.
The corresponding thermal reaction rate is approximately (here we use the Boltzmann distribution instead of Fermi-Dirac in order to obtain an analytic expression)
\begin{equation}
\Gamma_V^\text{on-shell} = \frac{|\lambda_{\mu\mu}|^2 m_V^2 T}{8\pi E^2} \left( \ln \left( 1 + e^{\omega} \right) - \omega \rule{0mm}{4mm}\right) \ ,
\end{equation}
where $w \equiv {m_V^2}/{(4ET)}$. Assuming $E\sim T$, compared to the $\Gamma^\text{high-T}$ in Eq.~(\ref{highTrate}), the on-shell rate has an extra factor $(m_V/T)^4$, which is more suppressed at high $T$.
This corresponds to a phase space (collinear) suppression for two energetic neutrinos to scatter at center-of-mass energy equal to $m_V$.
However, this factor is not important when the temperature drops to $T\sim m_V$. 
In addition, the on-shell rate only involves only two powers of the $|\lambda_{\mu\mu}|$ coupling, in contrast to the other rates, making it easier to stand out in the small coupling regime.

Numerically, we find that at $T\geq m_V$, $\Gamma_V$ is approximately given by $\Gamma^\text{on-shell}_{V} + \Gamma^\text{high-T}$, whereas at $T\leq m_V$
$\Gamma_V \simeq \Gamma^\text{on-shell}_{V} + \Gamma^\text{low-T}$.
These asymptotic rates in the two regimes capture the dominant contributions, and they match each other well at $T=m_V$ because the $\Gamma^\text{on-shell}_{V}$ term dominates, as long as $|\lambda_{\mu\mu}| <1$.

To summarize, in the presence of the $V$ boson mediated neutrino self interaction, the total thermal potential and reaction rate are given by 
\begin{equation}
V_T = V_{T,\,{\rm SM}}+V_{T,\, V}, \quad {\rm and}  \quad \Gamma = \Gamma_{\rm SM}+ \Gamma_V\ ,
\end{equation}
where $V_{T,\,{\rm SM}}$ and $\Gamma_{\rm SM}$ are the Standard Model contribution, given in Eqs.~(\ref{Eq:SM_VT}) and (\ref{Eq:SM_rate}), respectively. 
The new physics contribution $V_{T,\, V}$ is given by Eq.~(\ref{eq:VT}), and $\Gamma_V$ given by Eq.~(\ref{Eq:Gamma_def}). Note that the thermal potential is calculated at the amplitude level, so there will not be any interference terms while estimating the potential. Interference is, however, possible between the SM and the $V$-induced contributions to the thermal rates. However, since the the new interaction is considered to be much stronger than the ordinary weak interactions, the contribution of the interference terms in the thermal rates are sub-dominant, and hence neglected.
With the above potential and rate, we numerically solve the Boltzmann equation for the phase space distribution $f_{\nu_4}$, Eq.~\eqref{masterequation}, up to 
a sufficiently low temperature $T_f$ such that the value for $f_{\nu_4}$ saturates.
This temperature should be much lower compared to the mediator mass $m_V$ but still higher than the dark matter mass $m_4$. In practice, we take $T_f=100\,$keV. The dark matter relic density today corresponds to
\begin{equation}
\Omega \,\equiv\, \frac{Y_{\nu_4} s_0 m_4}{\rho_0},\quad  {\rm with\,\,\,} Y_{\nu_4}\, \equiv\, \frac{1}{s(T_f)}\int \frac{d^3\vec{p}}{(2\pi)^3} f_{\nu_4}(E,T_f) \ ,
\end{equation}
where $\rho_0 = 1.05\times 10^{-5} h^{-2}\,{\rm GeV/cm^3}$ is the critical density, $s$ is the entropy density as a function of temperature of the universe, and $s_0= 2891.2\,{\rm cm^{-3}}$ is the entropy density today. 

\begin{figure}[!t]
\centering\includegraphics[width=0.6\textwidth]{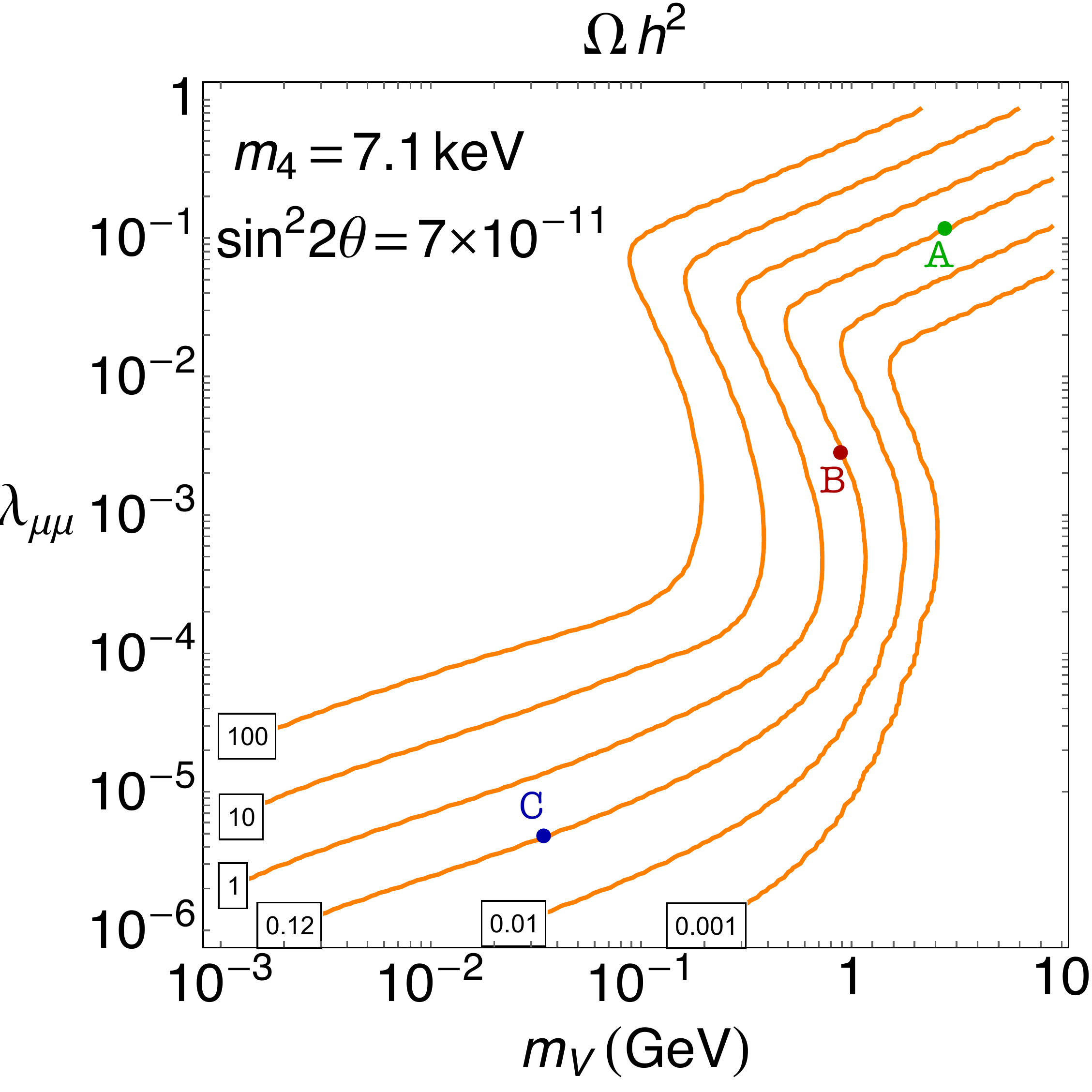}
\caption{Contours of $\Omega h^2$ for $ m_4= 7.1\,{\rm keV},\,\sin^2 2\theta=7\times 10^{-11}$, showing regions of over and under-abundance. The values of the relic density are labeled on the contours. Three benchmark points on the contour corresponding to the observed relic density are chosen:  A $(\lambda_{\mu\mu}=0.11,\, m_V=2.77\,{\rm GeV}) $, B $(\lambda_{\mu\mu}=0.003,\, m_V=0.88\,{\rm GeV}) $ and C $(\lambda_{\mu\mu}=4.8\times 10^{-6},\, m_V=0.03\,{\rm GeV})$. }
\label{fig:SterileDM_NuPhil_Scurve}
\end{figure}

The result of the numerical calculation described above is shown in Fig.~\ref{fig:SterileDM_NuPhil_Scurve}, in the parameter space of $\lambda_{\mu\mu}$ versus $m_V$, where the orange contours stand for constant values of the sterile neutrino dark matter relic density $\Omega h^2$. We choose a set of input parameters, $m_4= 7.1\,{\rm keV}$ and $\sin^2(2\theta)=7\times 10^{-11}$.  This point corresponds to the potential dark matter explanation of the $3.55\,{\rm keV}$ X-ray line. Interestingly, all the curves exhibit an ``$\mathcal{S}$-shape'', with three regimes having distinct slopes of the curve.
This diverse parametric dependence on $\lambda_{\mu\mu}$ and $m_V$ suggests different detailed production processes on which we shall elaborate on below. 
Along the curve where the correct relic density is reproduced ($\Omega h^2=0.12$), we mark three points A, B, C, one from each regime.

\begin{figure}[!t]
\centerline{\includegraphics[width=0.37\textwidth]{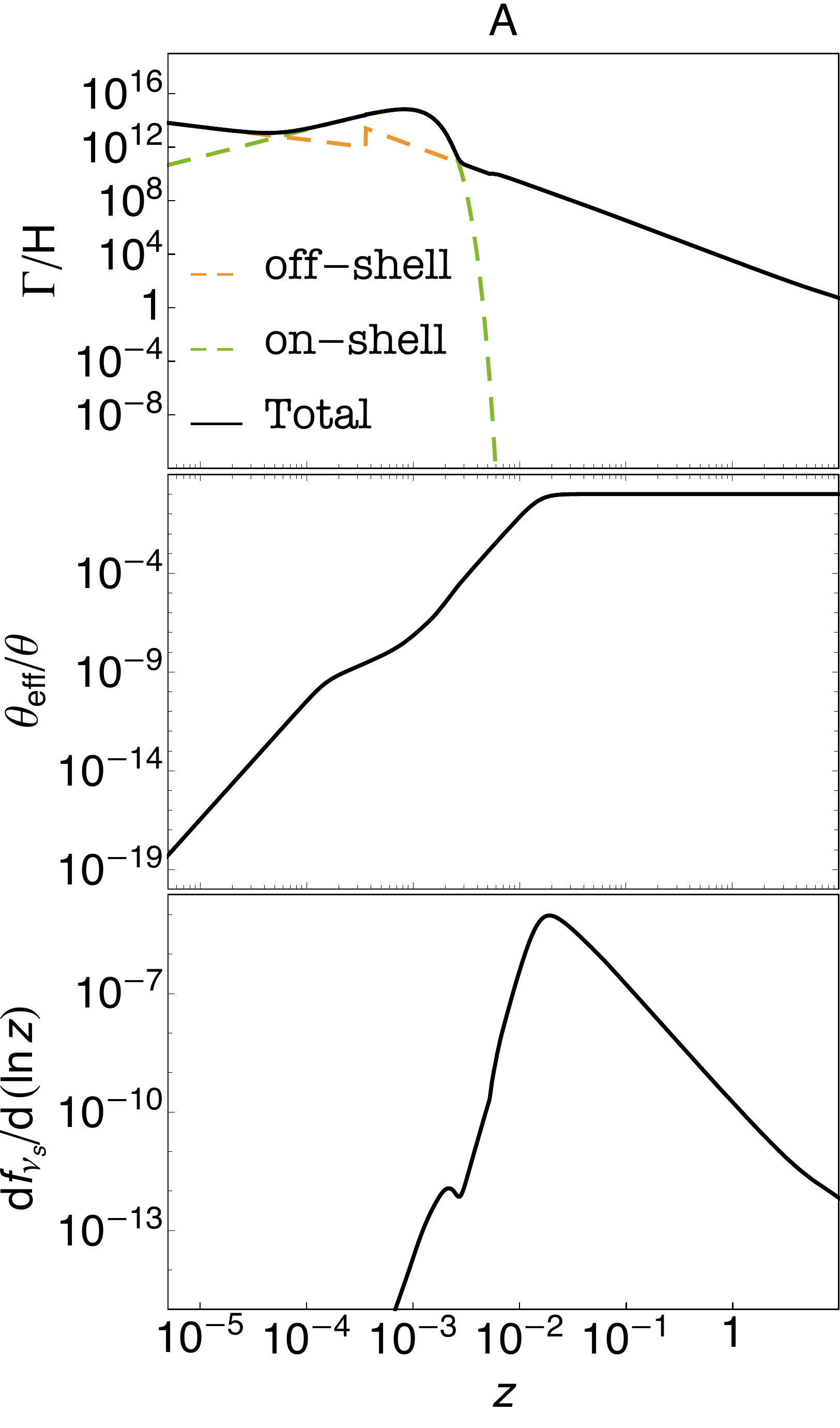}\hspace{-0.52in} \includegraphics[width=0.37\textwidth]{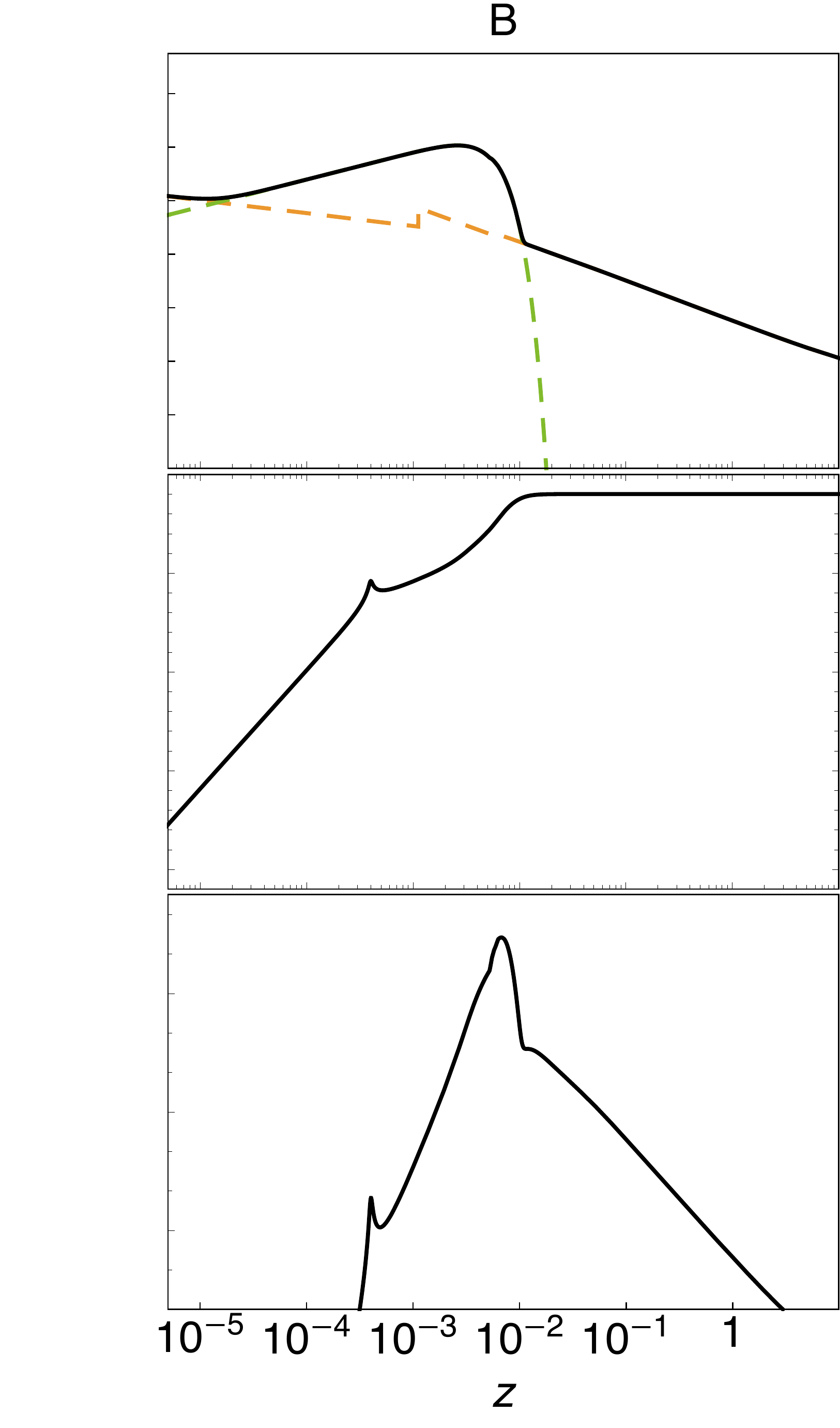}\hspace{-0.52in} \includegraphics[width=0.37\textwidth]{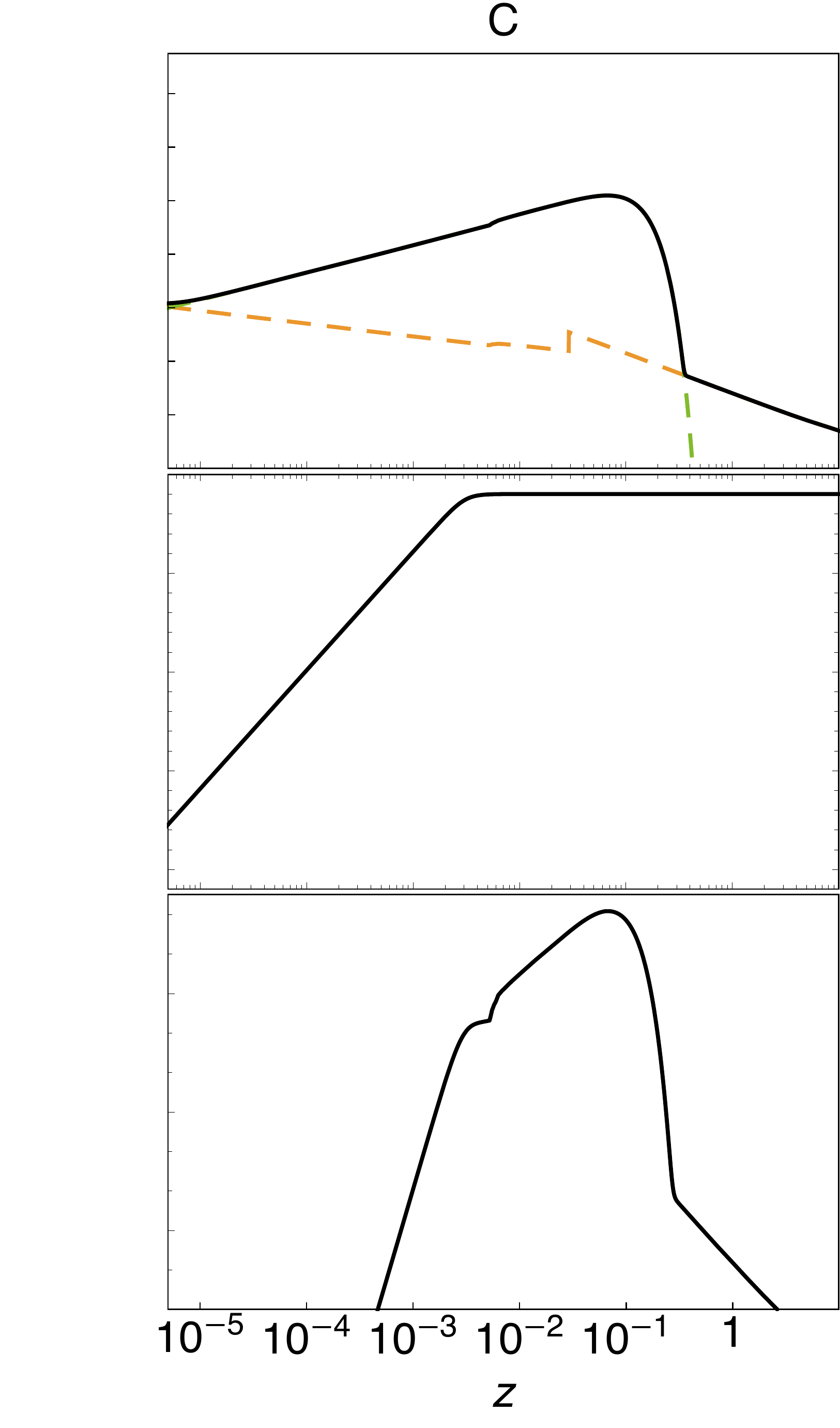}}
\caption{Plot showing the variation of the new thermal scattering rate as compared to the Hubble parameter (top), the ratio of the effective mixing angle to the vacuum angle (middle), and the differential sterile neutrino production rate (bottom)  as a function of $z$ for the three benchmark points A, B and C, as chosen in Fig.\,\ref{fig:SterileDM_NuPhil_Scurve}. 
}
\label{fig:ABC}
\end{figure}

The different parametric dependence mentioned above can be understood by examining Eq.~(\ref{masterequation}), which in the very small $\theta$ limit takes the form,
\begin{equation}
\frac{d\, f_{\nu_4}(x, z)}{d \ln z} \simeq \frac{\Gamma }{H} \,\theta_{eff}^2\,  f_{\nu}(x) 
\simeq \frac{\Gamma }{H} \,\frac{\Delta^2 \theta^2}{\Gamma^2/4 + (\Delta - V_T)^2}\,  f_{\nu}(x) \ .
\end{equation}
We plot different factors on the right-hand side of the above equation in Fig.~\ref{fig:ABC}. In the figure, each column corresponds to case A, B, or C, respectively. The first row depicts $\Gamma/H$ as a function of time (labeled by $z\equiv {\rm 1\,MeV}/T$) in each case. 
This ratio dictates the epoch when the active neutrino self interaction drops out of thermal equilibrium (when $\Gamma/H<1$).
In all cases the rate $\Gamma$ features a bump which corresponds to the exchange of an on-shell $V$ in the neutrino self interaction. This contribution is most important when $T\sim m_V$ but becomes Boltzmann suppressed at $T\ll m_V$. At low temperatures, the off-shell $V$ exchange dominates.

The second row of Fig.~\ref{fig:ABC} depicts the effective mixing angle $\theta_{eff}$ [defined in Eq.~(\ref{mixinganglematt})] divided by the vacuum mixing angle $\theta$. The value of $\theta_{eff}^2$ dictates the probability for each active neutrino in the thermal plasma to oscillation into a sterile neutrino. At very early time, $\theta_{eff}/\theta$ is highly suppressed because the oscillation frequency $\Delta$ is much smaller compared to $\Gamma$ and $V_T$. The production of the sterile neutrino state become more efficient at later time when $\theta_{eff}$ catches up with $\theta$.

The above temperature dependence lead to interesting interplay in the third row of Fig.~\ref{fig:ABC}, which depicts 
the actual dark matter production rate ${d\, f_{\nu_4}(x, z)}/{d \ln z}$ and is proportional to the product of $\Gamma/H$ and $\theta_{eff}^2$. 
Here we focus on the energy $E=T$ (or $x=1$) which has the highest population of active neutrinos as the source term.
We introduce three useful time scales:
{\bf i)} $z_0$: when $\Gamma/H=1$. At $z\gg z_0$, the active neutrino self interaction falls out of thermal equilibrium and dark matter production ceases;
{\bf ii)} $z_1$: when $\Delta \simeq {\rm Max}\{|V_T|, \Gamma_a\}$. At $z\ll z_1$, $\theta_{eff} \ll\theta$ implies less efficient production. At $z\gtrsim z_1$, $\theta_{eff} \simeq\theta$ implies efficient production;
and {\bf iii)} $z_2\equiv \mu/m_V$. At $z\gg z_2$, the vector boson $V$ is too heavy to be produced in the universe.

First of all, if $z_0$ is smaller than both $z_1$ and $z_2$, the new neutrino interaction via $V$ decouples too early and is irrelevant for dark matter production. We do not consider this case.
The three cases A, B, C mentioned above correspond to the following different hierarchies of $z_0, z_1, z_2$ and thus different sterile neutrino dark matter production scenarios.

\begin{itemize}

\item {\bf Case A} corresponds to heavy $V$ and large coupling $\lambda_{\mu\mu}$. In this case, the hierarchy of time scales satisfy, $z_2 < z_1 < z_0$. As a result, when the temperature is high enough to thermalize $V$, the mixing angle $\theta_{eff}$ is so suppressed that dark matter produced around this time ($z\sim z_2$) is negligible. The universe has to wait until a later time ($z\sim z_1$) for more efficient dark matter production.
Dark matter is mainly produced from active neutrino scattering via heavy off-shell $V$ exchange. 
We find $df_{\nu_s}/d{\rm ln}z \propto z^9$ for $z\lesssim z_1$, and $df_{\nu_s}/d{\rm ln}z \propto z^{-3}$ for $z\gtrsim z_1$. They lead to the triangle peak in the third row of case A in Fig.~\ref{fig:ABC} and the following parametric dependence in the final relic density,
\begin{equation}
\Omega_{\nu_4} \propto \frac{\theta^2 |\lambda_{\mu\mu}|^3 m_4^2}{m_V^2} \ .
\end{equation}
Case A also represents the standard Dodelson-Widrow mechanism with $|\lambda_{\mu\mu}|$ and $m_V$ replaced by the $SU(2)_L$ gauge coupling $g_2$ and the $W$-boson mass $M_W$ respectively.

\item {\bf Case B} has the hierarchy of time scales: $z_2 < z_0 < z_1$. The production of dark matter is restricted to the time window
$z_2 \lesssim z \lesssim z_0$, where the the effective mixing angle $\theta_{eff}$ is still suppressed compared to the vacuum angle $\theta$, but the universe remains hot enough compared to $V$ mass. Therefore, dark matter can be produced through active neutrino scattering via on-shell $V$ exchange, or equivalently, the decay (and inverse decay) of $V$ particles in the plasma.
In this case we find $df_{\nu_s}/d{\rm ln}z \propto z^7$ and the key parametric dependence in the final relic density is,
\begin{equation}
\Omega_{\nu_4} \propto \frac{\theta^2 m_4^5}{|\lambda_{\mu\mu}|^2 m_V^5} \ .
\end{equation}
This is a new production regime beyond the Dodelson-Widrow mechanism.

\item {\bf Case C} corresponds to a very light $V$ and tiny coupling $\lambda_{\mu\mu}$. In this case the time scales satisfy, $z_1 < z_2 < z_0$.  As a result, when the effective mixing angle $\theta_{eff}$ relaxes to the vacuum angle $\theta$, the universe is still hot enough compared to $V$ mass. Thus, dark matter is mainly produced from active neutrino scattering via on-shell $V$ exchange. In the dominant production window,  $z_1\lesssim z\lesssim z_2$, we find $df_{\nu_s}/d{\rm ln}z \propto z^3$. 
This implies the $z$ integral is dominated by $z \simeq z_2$, where $T \sim m_V$.
The key parametric dependence in the final relic density is,
\begin{equation}
\Omega_{\nu_4} \propto \frac{\theta^2 |\lambda_{\mu\mu}|^2 m_4}{m_V} \ .
\end{equation}
Like case B, this is another new production regime beyond Dodelson-Widrow.

\end{itemize}
In practice, we find that cases A, B, C cover all the possible production scenarios by varying the values of $m_{\nu_4}$ and $\theta$. 
The $\mathcal{S}$-shape of the relic density contours is an interpolation through the three cases.
In case A, the dominant production temperature is $T\ll m_V$, whereas in cases B and C, the dominant production temperature is $T\sim m_V$.
Thanks to this, thermal corrections to the $V$ mass has little impact on the result.
Furthermore, in both cases B and C where the $V$ boson is light enough to be on-shell, the values of $\lambda_{\mu\mu}$ for relic density is always much smaller than 1. 
This justifies the approximation made that neglects the scattering channel, $\nu + V \to \nu +V$, whose rate is higher order in $\lambda_{\mu\mu}$.

Mathematically, another hierarchy $z_1<z_0<z_2$ is also possible. However, for the phenomenologically allowed values of $\lambda_{\mu\mu}$ and $m_V$ considered in this work, there is no room for $z_0<z_2$ to occur, i.e., the $V$ boson always becomes heavy before the new interaction decouples.

\begin{figure}[t]
\centering\includegraphics[width=0.72\textwidth]{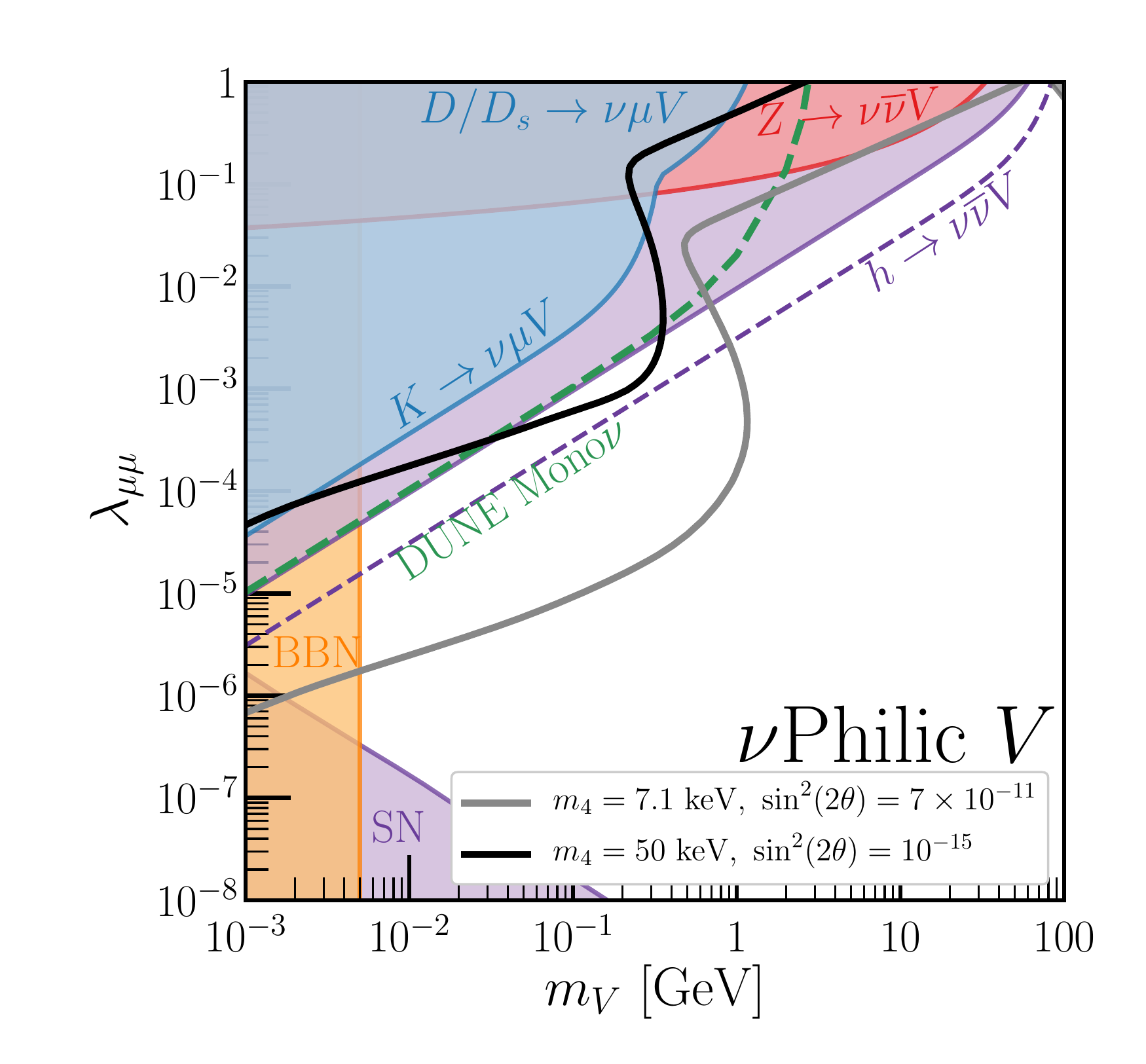}
\caption{Curves on the parameter space that yield the observed relic density of dark matter today for the neutrinophilic vector boson model and $ m_4= 7.1\,{\rm keV},\,\sin^2(2\theta)=7\times 10^{-11}$ (gray), and  $m_4= 50\,{\rm keV},\,\sin^2(2\theta)=10^{-15}$ (black). Existing limits from a variety of probes are shown filled, colored regions: rare meson decays (blue), invisible $Z$ boson decay (red), invisible Higgs boson decay (purple), BBN (orange), and Supernova 1987A (purple). Future constraints from DUNE (green) and the high-luminosity LHC for invisible Higgs boson decay (purple) are shown as dashed lines.}
\label{fig:SterileDM_NuPhil}
\end{figure}

For varying values of dark matter mass $m_4$ and mixing angle $\theta$, the $\mathcal{S}$-shape relic density curve sweeps across the $\lambda_{\mu\mu}$ versus $m_V$ plane.
In Fig.~\ref{fig:SterileDM_NuPhil}, we plot two choices of parameters:
\begin{itemize}
\item Gray curve: $m_4= 7.1\,{\rm keV}$, $\sin^2(2\theta)=7\times 10^{-11}$.
\item Black curve: $m_4= 50\,{\rm keV}$, $\sin^2(2\theta)=10^{-15}$.
\end{itemize}
For heavier sterile neutrino dark matter (black curve), we choose a correspondingly smaller mixing angle in order to satisfy the X-ray constraints.
We find that heavier dark matter generically requires larger $|\lambda_{\mu\mu}|$ and/or smaller $m_V$ for the relic density, making it more constrained experimentally.
On the other hand, due to the small-scale structure bounds arising out of free-streaming considerations of the sterile neutrinos as well as 
dwarf galaxy and Lyman-$\alpha$ constraints, there is limited room to make the dark matter $\nu_4$ much lighter than 7.1 keV. We have explicitly checked this for the three benchmark points considered above.
As a result, the gray curve in Fig.~\ref{fig:SterileDM_NuPhil} roughly indicates the lower margin of parameter space where $\nu_4$ comprises the total dark matter relic density in the universe. 
I.e., all the target parameter space lies above the gray curve.

The existence of $V$ also mediates decays of the type $\nu_4 \to \nu\nu\bar{\nu}$ via an off-shell $V$. This decay width is proportional to $\lambda_{\mu\mu}^4 \sin^2(2\theta) m_4^5/m_V^4$. If we want $\nu_4$ to be cosmologically long-lived and its lifetime to be longer than the age of the universe, this provides an indirect constraint on the model parameters. We find that, for the masses $m_4$ and mixings $\sin^2(2\theta)$ that we consider, this constraint is weaker than the other bounds to be discussed in the following subsections.

\subsection{Constraint from the Higgs boson invisible decay}\label{sec:HiggsInvis}

In this and the next few subsections, we present experimental constraints on the neutrinophilic $V$ boson.
To obtain the correct dark matter relic density, the mass of $V$ is required to be smaller than the weak scale. 
It mainly decays into neutrinos and thus appears invisible after production in laboratories.
This leads to a number of constraints by making precision measurements of processes that involve an active neutrino.

The first process we consider is the Higgs boson decay. The operator introduced in Eq.~(\ref{eq:Phen}) opens up a new Higgs invisible decay channel, $h\to \nu_\mu\bar{\nu}_\mu V$. The corresponding partial decay rate is
\begin{equation}
\Gamma(h\to \nu_\mu\bar{\nu}_\mu V) = \frac{\left\lvert\lambda_{\mu\mu}\right\rvert^2 G_F^2 M_H^5}{3072\sqrt{2}\pi^3 m_V^2} \left(1 + 12h_V^2\left(6 + (3+4h_V)\log{(h_V)}\right) - 64h_V^3 - 9h_V^4\right),
\end{equation}
where $h_V \equiv m_V^2/m_h^2$. We see that in the small $m_V$ limit, this partial width is enhanced by a factor of $1/m_V^2$, corresponding to the Higgs decaying into the longitudinal component of the $V$ boson. The operator responsible for Higgs decay, $\lambda (h/v) \bar \nu \gamma^\mu \nu V_\mu$, does not admit any conserved $U(1)'$ gauge symmetry for $V^\mu$ to be promoted as the corresponding gauge boson.
For example, the $U(1)$ of neutrino number is explicitly broken by the presence of the Higgs field.

To set a limit using this channel, we note that the invisible decay branching ratio of the Higgs boson in this model is calculated as, $\Gamma_{h\to \nu_\mu\bar{\nu}_\mu V}/\Gamma_{\rm total}$,
where the Higgs total width $\Gamma_{\rm total}$ is the sum of the Standard Model Higgs width ($\sim 4\,$MeV) and $\Gamma_{h\to \nu_\mu\bar{\nu}_\mu V}$. The present LHC upper bound on Higgs invisible branching ratio, ${\rm Br}(H\to {\rm invisible}) < 24\%$~\cite{Tanabashi:2018oca}, leads to a constraint in the $\lambda_{\mu\mu}$ versus $m_V$ parameter space, corresponding to the purple shaded region in Fig.~\ref{fig:SterileDM_NuPhil}. 
The future running of high-luminosity (HL) LHC is expected to further improve the above limit to ${\rm Br}(H\to {\rm invisible})< 2.5\%$~\cite{Cepeda:2019klc}, which corresponds to the purple dashed curve in Fig.~\ref{fig:SterileDM_NuPhil}.
We find this is the leading constraint on the model parameter space for $m_V>5\,$MeV.
In particular, for the case $m_4= 50\,{\rm keV}$ and $\sin^2(2\theta)=10^{-15}$, the present Higgs invisible limit still allows a region with $m_\phi\sim100$\,MeV, but that will be covered by the HL-LHC.
It is also worth pointing out that the Higgs invisible decay constraint equally applies to $V$ coupling to other neutrino flavors.

\subsection{Constraint from $W$ boson decay width}

The next channel we examine is the $W^- \to \mu^-\bar \nu_\mu  V$ where $V$ is radiated from the final state $\bar\nu_\mu$ via its neutrinophilic coupling. Ignoring the muon mass, this decay rate is
\begin{equation}
\Gamma(W^-\to \mu^- \bar{\nu}_\mu V) = \frac{\left\lvert \lambda_{\mu\mu}\right\rvert^2 G_F^2 M_W^5}{512\sqrt{2} \pi^3 m_V^2} \left( 1 - w_V - 12 w_V^2\log{(w_V)} + 8w_V^3 - w_V^4\right),
\end{equation}
where $w_V \equiv m_V^2/M_W^2$. As with the Higgs decay, we see that in the small $m_V$ limit, this decay width is  enhanced by $1/m_V^2$. To derive a conservative constraint, we simply require this decay rate to be smaller than the uncertainties in the $W$ total width measurement,  $\sim 42$\,MeV~\cite{Tanabashi:2018oca}. The resulting limit is found to be much weaker than the one from Higgs invisible decay, 
and we do not show it in Fig.~\ref{fig:SterileDM_NuPhil}. Including the muon mass in this calculation would drive the constraint to be slightly weaker than this calculation.

We note that the $W$ decay limit might be further improved by studying the kinematic edge observables of $W \to \mu + {\rm MET}$ decay at the LHC. In the Standard Model, when the $W$-boson is singly produced, the final charged lepton transverse momentum distribution features a Jacobian peak~\cite{Baak:2013fwa}. This feature is absent when $V$ is present in the decay product.
We leave a careful study of this for a future work.

\subsection{Constraint from $Z$ boson invisible width}

We also examine the $Z \to \nu_\mu \bar \nu_\mu V$ decay where $V$ is radiated from either $\nu_\mu$ or $\bar\nu_\mu$ in the the final state. 
Because $V$ decays into neutrino and appears invisible at colliders, this decay channel contributes to additional invisible $Z$-boson decay width.
This decay rate is
\begin{eqnarray}\label{eq:ZDecay}
\begin{split}
\Gamma(Z \to \nu_\mu \bar \nu_\mu V) &= \frac{\left\lvert \lambda_{\mu\mu}\right\rvert^2G_F M_Z^3}{48\sqrt{2} \pi^3} \int_{0}^{\left( 1 -{m_V}/{M_Z}\right)^2} g_Z \left({m_V^2}/{M_Z^2},  w\right) dw \ . \\
g_Z\left(y, w\right) &=  \frac{(y+1)^2+w^2}{y-w+1} \log \left[ \frac{y-w+1+\sqrt{w^2 - 2 w (y+1) + (y-1)^2}}{y-w+1-\sqrt{w^2 - 2 w (y+1) + (y-1)^2}} \right]  \\
& \quad -2\sqrt{y^2 - 2y(w+1)+(w-1)^2} \ .
\end{split}
\end{eqnarray}
Unlike the Higgs and $W$ cases, the new $Z$ boson decay width does not feature the $1/m_V^2$ enhancement at small $m_V$. 
This could be understood as follows.
In this decay process, the vector boson $V$ couples to $\bar\nu_\mu\gamma^\mu \nu_\mu$ which can be defined as the neutrino current. The $U(1)'$ symmetry for $\nu_\mu$ neutrino number associated with this current is preserved by the Standard Model $Z\nu_\mu\bar\nu_\mu$ coupling.
This symmetry is only broken by the other Standard Model couplings ({\it e.g.}, the $W\mu \bar\nu$ coupling considered earlier) which do not take part in this decay rate at leading order.
Due to the lack of the $1/m_V^2$ factor, numerically, we also find the constraint from $Z$ invisible decay is much weaker than that from Higgs invisible decay derived above. It is shown as the red shaded region in Fig.~\ref{fig:SterileDM_NuPhil}.

\subsection{Constraints from exotic meson decays}

For the vector boson $V^\mu$ lighter than GeV scale, the $V\nu\bar\nu$ coupling could also lead to new decay channels of charged mesons will exist, $\mathfrak{m^+} \to \mu^+ \nu_\mu V$, where $\mathfrak{m}= \pi, K, B$, etc. Because $V$ will invisibly decay back to neutrinos, such processes would contribute to the branching ratios of the $\mathfrak{m^+} \to \mu^+ \nu_\mu \nu \bar{\nu}$ and get constrained.
The decay width $\mathfrak{m}^+ \to \ell^+ \nu V$ is
\begin{eqnarray}\label{eq:MesonDecay}
\begin{split}
&\Gamma(\mathfrak{m}^+ \to \mu^+ \nu_\mu V) = \frac{\left\lvert \lambda_{\mu\mu} V_{q q^\prime}\right\rvert^2G_F^2 F_\mathfrak{m}^2 m_\mathfrak{m}^5}{64\pi^3 m_V^2} \int_{m_V^2/m_\mathfrak{m}^2}^{\left( 1 - {m_\ell}/{m_V}\right)^2} f_\mathfrak{m} \left({m_\ell^2}/{m_\mathfrak{m}^2}, {m_V^2}/{m_\mathfrak{m}^2}, w\right) dw, \\
&f_\mathfrak{m}\left(l,v,w\right) = \frac{1}{w^3} \sqrt{l^2 - 2l(w+1) + (1-w)^2} \left[-l^2 + l(2w+1) + w(1-w)\right](v-w)^2 (2v+w),
\end{split}
\end{eqnarray}
where $G_F$ is the Fermi constant, $F_\mathfrak{m}$ is the decay constant of $\mathfrak{m}$, and $|V_{q q^\prime}|$ is the relevant CKM matrix element for this decay. 
In the small $m_V$ limit, this integral may be performed analytically,
\begin{equation}
\Gamma(\mathfrak{m}^+ \to \mu^+ \nu_\mu V) \underset{m_{\ell}\to0}{\longrightarrow} \frac{\left\vert\lambda_{\mu\mu} V_{q q^\prime}\right\rvert^2 G_F^2 F_\mathfrak{m}^2 M_\mathfrak{m}^5}{768\pi^3 m_V^2} \left(1 + \frac{m_V^2}{M_\mathfrak{m}^2} + 73 \frac{m_V^4}{M_\mathfrak{m}^4} + 9\frac{m_V^6}{M_\mathfrak{m}^6}\right).
\end{equation}

In practice, we have considered the charged kaon and pion decays, and apply the experimental constraints, ${\rm Br}(K^+\to \mu^+\nu_\mu\nu\bar\nu) < 2.4\times 10^{-6}$ and ${\rm Br}(\pi^+\to \mu^+\nu_\mu\nu\bar\nu) < 5\times 10^{-6}$~\cite{Tanabashi:2018oca}.
The resulting limit is shown by the blue shaded region in Fig.~\ref{fig:SterileDM_NuPhil}.
It is weaker than that from Higgs invisible decay.

Searches for exotic lepton decays can provide similar constraints, however, they are suppressed by a larger final-state phase space, e.g. the decay $\mu^- \to e^- \nu\overline{\nu} V$. Because the kaon decay constraints are strong in this region of parameter space, we expect that this four-body phase space would lead to relatively weaker constraints than those of the meson decays, which are, in turn, weaker than the Higgs boson decay constraints (the Higgs decay constraint also benefits from the longitudinal enhancement, with a partial width that scales like $1/m_V^2$). For heavier $V$, we could use the decay $\tau^- \to \ell \nu \overline{\nu} V$, which suffers from the same four-body suppression. This constraint has been calculated in the case of a new-physics scalar instead of a vector in Ref.~\cite{Brdar:2020nbj}, where it was found to be much weaker than other laboratory based probes. We expect the same to be true for a vector $V$ being emitted in this decay.

\subsection{Constraints from BBN and Supernova 1987A}

If the new vector boson $V$ is thermalized in the early universe and light enough to remain relativistic by the time of big bang nucleosynthesis, it will contribute to $\Delta N_{\rm eff.}$ and affect the primordial element abundances. A conservative constraint rules out $m_V \lesssim 5$ MeV~\cite{Blinov:2019gcj}. For a detailed computation of the effects of non-standard neutrino self-interactions on big bang nucleosynthesis, see~\cite{Grohs:2020xxd}.

If the vector $V$ is lighter than $\sim100\,$MeV, it can be produced from neutrino scatterings in the explosion of Supernova 1987A. It could carry away significant fraction of energy and modify the observed time scale of neutrino emission. Ref.~\cite{Escudero:2019gzq} has estimated this effect in the gauged $U(1)_{L_\mu - L_\tau}$ model (see also Section~\ref{sec:LmuLtau}). We rescale and apply their bound by restricting $V$ to couple to only one neutrino flavor.

\subsection{Mono-neutrino probes at DUNE}\label{sec:DUNE}

Neutrinophilic particles like $V$ may be emitted in beam neutrino experiments, if kinematically allowed. The processes $\nu_\mu n \to V \mu^- p^+$ occurs via initial state radiation. In contrast to standard neutrino quasi-elastic scattering, $\nu_\mu n \to \mu^- p^+$,
the radiation of $V$ carries away both energy and transverse momentum.
As a result, the final state $\mu^-$ carries lower energy than expected in the quasi-elastic scattering case.
The resulting $\mu^- p^+$ system appears to have a non-zero $p_T$ with respect to the beam direction.
These are the mono-neutrino signals introduced in~\cite{Kelly:2019wow, Berryman:2018ogk} in a different context.
In contrast to the lepton-number charged scalar radiation studied carefully in~\cite{Kelly:2019wow}, here the radiation of $V$ does not carry away lepton-number. The final state muon in the above signal process has the same electric charge as the quasi-elastic background, thus it will not benefit form the muon charge identification capability of the neutrino detector. In contrast, the emission cross section for a light $V$ features a $E^2/m_V^2$ longitudinal enhancement, making the signal stronger at low $m_V$. 
We derive an expected 95\% CL constraint assuming five years of DUNE~\cite{Abi:2020evt} data collection in the neutrino mode, as shown by the dashed green line in Fig.~\ref{fig:SterileDM_NuPhil}. An additional five years of DUNE data collection in antineutrino mode does not improve this limit significantly, as the background processes are more difficult to resolve when an antineutrino scatters. Our obtained sensitivity is comparable to the current Higgs invisible limit but will be surpassed by the high-luminosity LHC search.

\subsection{On $V$ coupling to other neutrino flavors}\label{subsec:DifferentFlavors}

Throughout Section~\ref{sec:neutrinphilic} we have focused on $\lambda_{\mu\mu}$, the $V$ coupling to muon-flavored neutrinos. In principle, the couplings $\lambda_{\alpha\beta}$ comprise a $3\times3$ matrix including couplings to other flavors and even flavor-violating couplings. Since all of the early-universe reactions that lead to the production of $\nu_4$ are driven by the self interaction of neutrinos, the relic density results shown in Figs.~\ref{fig:SterileDM_NuPhil_Scurve} and \ref{fig:SterileDM_NuPhil} are independent of the choice of flavor, as long as one only element of $\lambda$ is turned on each time. 
Experimentally, the strongest constraint on $\lambda_{\mu\mu}$ comes from the Higgs boson invisible decay, as shown in Fig.~\ref{fig:SterileDM_NuPhil}. It equally applies to all other $\lambda_{\alpha\beta}$ couplings.

On the other hand, we note that if the $V$ coupling is flavor off diagonal, there will be strong constraints from loop induced $\mu\to e V$, $\tau\to e V$, $\tau\to \mu V$ decays. Given order-of-magnitude estimates of this loop process, we find that these constraints would be stronger than all of those discussed above and would rule out the desired parameter space for relic sterile neutrino dark matter.

\subsection{A possible UV completion}\label{subsec:NeutrinophilicUV}

In this subsection, we present a simple UV completion for the operator introduced in Eq.~(\ref{eq:Phen}). It serves as a proof-of existence of renormalizable theories that could lead to the low energy effective theory considered in this work.

We extend the Standard Model with a pair of chiral fermions $N_L$, $N_R$ and a complex scalar $\phi$. All are Standard Model gauge singlets. The $N_L$ and $\phi$ fields are oppositely charged under a new $U(1)'$ gauge symmetry whereas $N_R$ is neutral. With such a particle content the $U(1)'$ still possess a gauge anomaly that could be canceled by resorting to including additional heavy fermions without direct couplings to the lighter fields. We will keep those heavy particles implicit.

With the $N_L$, $N_R$ and $\phi$ fields, we could write down the following renormalizable Lagrangian governing their interactions,
\begin{eqnarray}\label{eq:LUV}
\mathcal{L}_{\rm UV} = (D_\mu \phi)^\dagger (D_\mu \phi) + \bar N_L i \cancel{D} N_L + \bar N_R i \cancel{\partial} N_R  + \left[ y \bar N_R L H + \lambda \bar N_R N_L \phi + {\rm h.c.} \rule{0mm}{4mm}\right] + V(\phi) \ , \nonumber \\
\end{eqnarray}
where $L, H$ are the Standard Model lepton and Higgs doublets, and $D_\mu = \partial_\mu \pm i g' V_\mu$.

We assume $\phi$ has a potential $V(\phi)$ such that $\phi$ gets a vacuum expectation value $v_\phi$ and breaks the $U(1)$ symmetry. This gives a mass to the new gauge boson $V$,
\begin{eqnarray}
M_{V} = \sqrt2 g' v_\phi \ ,
\end{eqnarray}
and also give a Dirac mass for the $N_L, N_R$ fermions (preferably above the electroweak scale for the discussions of constraints in previous subsections to be valid),
\begin{eqnarray}
M_N = \lambda v_\phi \ .
\end{eqnarray}
Next, we go to energy scales below $M_N$ and integrate out $N_L, N_R$. From the square bracket in Eq.~(\ref{eq:LUV}) the equation of motion of $\bar N_R$ yields (neglecting its kinetic term)
\begin{eqnarray}
N_L = - \frac{1}{M_N} y L H \ .
\end{eqnarray}
As a result, the gauge interaction of $N_L$ with the $V$ in the first line of Eq.~(\ref{eq:LUV}) leads to the effective interaction
\begin{eqnarray}
\mathcal{L}_{\rm eff} = \frac{g' y^2}{M_N^2} V_\mu (\bar L H^\dagger ) \gamma^\mu (LH) \ .
\end{eqnarray}
This is same as the higher dimensional operator introduced in Eq.~(\ref{eq:Phen}), with $\Lambda^2 = M_N^2/(g' y^2)$.
The presence of the new Dirac fermion $N$ in this UV completion could be subject to further experimental constraints (see e.g. \cite{deGouvea:2015euy}). 

As the final remark, it is worth mentioning that in $\mathcal{L}_{\rm UV}$ we suppress the renormalizable couplings of $N_{L}\phi$, $N_R$ with the sterile neutrino field $\nu_s$ in order to maintain the lightness of dark matter candidate.
We also neglect the Majorana mass term for $N_R$, which if exists, will contribute to the active neutrino mass via the inverted seesaw mechanism.
Again, we emphasize that $\mathcal{L}_{\rm UV}$ serves the role as proof-of existence that the effective operator Eq.~(\ref{eq:Phen}) can be derived from a renormalizable theory at higher scales.
We leave the detailed study of its phenomenology for a future work.

\section{Gauged $U(1)_{L_\mu-L_\tau}$ Model}\label{sec:LmuLtau}

The second model we explore is an extension of the Standard Model with gauged $U(1)_{L_\mu-L_\tau}$ symmetry.  It is one of the simplest anomaly-free $U(1)$ models known to provide an explanation to the discrepancy between the Standard Model prediction of muon $g-2$ and the experimental measurement~\cite{Baek:2001kca}. The phenomenology of this model has been explored extensively~\cite{Ma:2001md, Gninenko:2001hx, Heeck:2011wj, Harigaya:2013twa, Altmannshofer:2014pba, Kamada:2015era,Arcadi:2018tly}.
Here, we point out that the model also contains a target parameter space for sterile neutrino dark matter relic density.

The couplings of the $U(1)_{L_\mu-L_\tau}$ gauge boson $V$ with Standard Model fermions are
\begin{equation}
g_{\mu\tau} V_\alpha (\bar \mu \gamma^\alpha \mu - \bar \tau \gamma^\alpha \tau + \bar \nu_\mu \gamma^\alpha P_L \nu_\mu - \bar \nu_\tau \gamma^\alpha P_L \nu_\tau ) \ ,
\end{equation}
where $g_{\mu\tau}$ is the gauge coupling. 

Like the previous section, we assume that the sterile neutrino dark matter $\nu_4$ has a small $\nu_\mu$ component, characterized by a mixing angle $\theta$. To generate such a mixing, one cannot rely on the Yukawa coupling in Eq.~(\ref{eq:Sterile}). Instead, the following higher-dimensional operator must be introduced,
\begin{equation}\label{eq:Sterile2}
y_\varepsilon  \nu_s^T C H^T (i\sigma_2) L \left(\frac{\phi}{\Lambda} \right)  + {\rm h.c.}  \ ,
\end{equation}
where $\phi$ is a complex scalar field that Higgses the $U(1)_{L_\mu-L_\tau}$ symmetry and carries an opposite charge to that of $L = (\nu_\mu, \mu)^T$. This is because the $\nu_s$ field $(\simeq \nu_4)$ must be a singlet under $U(1)_{L_\mu-L_\tau}$ in order to prevent dark matter from being fully thermalized in the early universe and violating the initial condition for the production mechanism discussed in this work.

The relic density calculation in this model proceeds in similar fashion as the neutrinophilic one in Sec.~\ref{sec:neutrinphilic}.
The contribution to the thermal potential $V_T$ is the same as Fig.~\ref{fig:neutrinophilicV} with the gauge boson $V$ and $\nu$ running in the loop.
It is calculated using Eq.~(\ref{eq:VT}) with $\lambda_{\mu\mu}$ replaced by the gauge coupling $g_{\mu\tau}$.
In contrast, the thermal reaction rate of neutrino receives contributions from the scattering with
$\nu_\mu, \bar\nu_\mu, \nu_\tau, \bar\nu_\tau, \mu^\pm, \tau^\pm$ particles, via the new $L_\mu-L_\tau$ gauge interaction.

To simplify the thermal averaging in the numerical calculation, we make the instantaneous decoupling approximation.
For temperatures above the mass of a fermion participating in a scattering process, we calculate the thermal rates assuming the fermion to be massless. 
In contrast, when the temperature drops below any of the fermion mass, we assume the process vanishes immediately. 
Because sterile neutrino dark matter is dominantly produced at temperatures below a few hundred MeV (see, e.g., Fig.~\ref{fig:ABC} in  subsection~\ref{sec:anatomy}), 
the $\tau^\pm$ leptons are already decoupled from the universe, and the presence of muons is marginal.
This is a good approximation to simplify the numerical workload.

Under this approximation, we report the  cross sections and thermal rates relevant to the relic density calculation in this model:
\begin{equation}
\begin{split}
\sigma_{\nu_\mu \nu_\mu \to \nu_\mu \nu_\mu} &= \frac{g_{\mu\tau}^4}{4\pi m_V^2} 
\left[ \frac{s}{s+m_V^2} + \frac{2 m_V^2}{s + 2 m_V^2} \log\left( 1 + \frac{s}{m_V^2} \right) \right] \ , \\
\sigma_{\nu_\mu \bar \nu_\mu \to \nu_\mu \bar \nu_\mu} &= \frac{g_{\mu\tau}^4}{4\pi (s - m_V^2)^2}
\left[ \frac{s^2}{m_V^2} - 4 m_V^2 + \frac{4 \left( m_V^4 - s^2 \right)}{s}  \log\left( 1 + \frac{s}{m_V^2} \right)
+ \frac{10 s}{3} \right] \ , \\
\sigma_{\nu_\mu \bar \nu_\mu \to \nu_\tau \bar \nu_\tau} &= \frac{g_{\mu\tau}^4 s}{12\pi (s - m_V^2)^2} \ , \\
\sigma_{\nu_\mu\nu_\tau\to\nu_\mu\nu_\tau} &= \sigma_{\nu_\mu\bar\nu_\tau\to\nu_\mu\bar\nu_\tau} = \frac{g_{\mu\tau}^4 s}{4\pi m_V^2 (s + m_V^2)} \ , \\
\sigma_{\nu_\mu\bar\nu_\mu\to\mu^+\mu^-} &= \frac{g_{\mu\tau}^4 s}{6\pi (s-m_V^2)^2} \ , \\
\sigma_{\nu_\mu\mu^-\to\nu_\mu\mu^-} &= \sigma_{\nu_\mu\mu^+\to\nu_\mu\mu^+}
= \frac{g_{\mu\tau}^4 }{4\pi s^2} \left[ (s+m_V^2) \log \frac{m_V^2}{s+m_V^2} + 
\frac{s (2s^2 + 3 s m_V^2 + 2 m_V^4)}{2m_V^2 (s + m_V^2)} \right] \ .
\end{split}
\end{equation}
From the experience gained in the previous section (see Fig.~\ref{fig:ABC}), the most relevant neutrino reactions for dark matter production
occur at $T\lesssim m_V$. We present the thermal rate results for the heavy $V$ and on-shell $V$ cases.
In the heavy-$V$ limit ($m_V \gg T$),
\begin{equation}
\Gamma_V \approx \frac{7 \pi g_{\mu\tau}^4}{540 m_V^4} E \,T^4 \times
\left\{ 
\begin{array}{cl}
37,& \hspace{1cm} T\geq 2m_\tau \\
33,& \hspace{1cm} m_\tau \leq T < 2m_\tau \\
27,&\hspace{1cm} 2m_\mu \leq T< m_\tau \\
25,&\hspace{1cm} m_\mu \leq T< 2m_\mu \\
17,& \hspace{1cm} T<m_\mu \, .
\end{array}\right.
\end{equation}
For on-shell $V$ contribution ($m_V \lesssim T$), the corresponding rate is
\begin{equation}
\Gamma_{V} \simeq \frac{g_{\mu\tau}^4 m_V^2 T}{48\pi^2 \gamma_V E^2}\left( \ln \left( 1 + e^{\omega} \right) - \omega \rule{0mm}{4mm}\right) \,\times
\left\{ 
\begin{array}{cl} 
3,& \hspace{1cm} T\geq 2m_\tau \\
2,&\hspace{1cm} 2m_\mu\leq T< 2m_\tau \\
1,& \hspace{1cm} T<2m_\mu
\end{array}\right.
\end{equation}
where $w \equiv {m_V^2}/{(4ET)}$. $\gamma_V$ is the decay width of the $L_\mu-L_\tau$ gauge boson,
\begin{equation}
\gamma_{V} = \frac{g_{\mu\tau}^2 m_V}{12\pi} \left[ 1 + \sum_{\alpha=\mu}^\tau (1 + 2 r_\alpha) (1-4r_\alpha)^{1/2} \Theta (1-4r_\alpha) \right] \ ,
\end{equation}
where $r_\alpha = m_\alpha^2/m_V^2$ and $\Theta$ is the Heaviside theta function.

\begin{figure}[t]
\centering\includegraphics[width=0.72\textwidth]{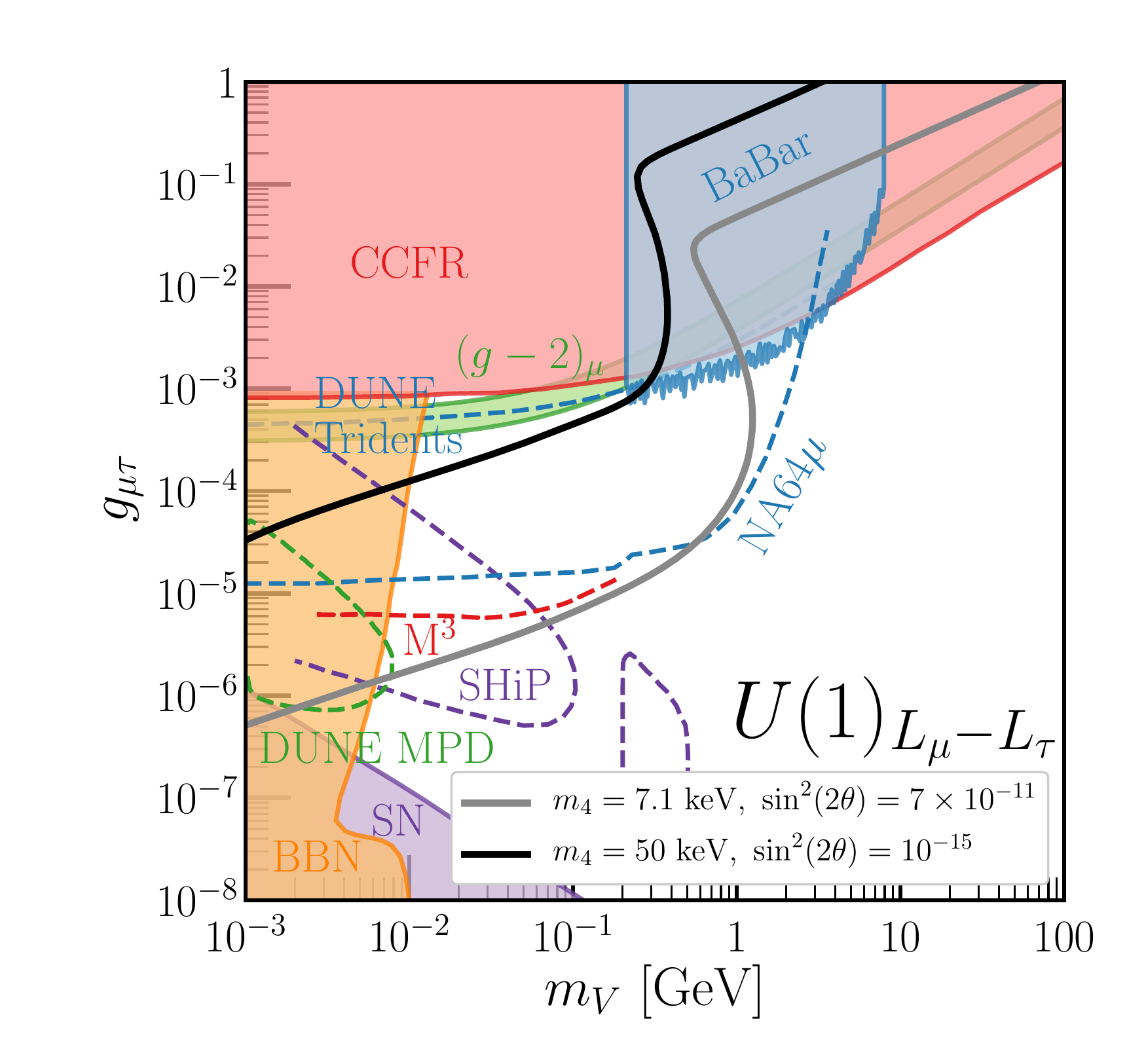}
\caption{Curves on the parameter space that yield the observed relic density of dark matter today for the $L_\mu - L_\tau$ vector boson model and $ m_4= 7.1\,{\rm keV},\,\sin^2(2\theta)=7\times 10^{-11}$ (gray), and  $m_4= 50\,{\rm keV},\,\sin^2(2\theta)=10^{-15}$ (black). Existing limits from a variety of probes are shown as shaded regions: neutrino trident scattering with CCFR (red), BaBar (blue), BBN (orange), and Supernova 1987A (purple). The green filled region corresponds to the preferred region for the $(g-2)_\mu$ anomaly. Future constraints from NA62 (red), NA64$\mu$ (blue), SHiP (purple), M$^3$ (red), and DUNE (green for MPD decays and blue for Trident scattering) are shown as dashed lines.}
\label{fig:SterileDM_Lmu-Ltau}
\end{figure} 

The above thermal potential and reaction rates are added to their counterparts in the Standard Model and then inserted to the Boltzmann equation (\ref{masterequation}).
The dark matter relic density favored parameter space is shown in the $g_{\mu\tau}$ versus $m_V$ plane in Fig.~\ref{fig:SterileDM_Lmu-Ltau}.
Similar to Fig.~\ref{fig:SterileDM_NuPhil}, the gray and black curves corresponds to two choices of parameters: 
$m_4= 7.1\,{\rm keV}$, $\sin^2(2\theta)=7\times 10^{-11}$ and
$m_4= 50\,{\rm keV}$, $\sin^2(2\theta)=10^{-15}$, respectively.
The relic density curves exhibit similar $\mathcal{S}$-shapes, corresponding to the three production regimes discussed in subsection~\ref{sec:anatomy}.

Sterile neutrino dark matter production in this model has been explored in~\cite{Shuve:2014doa} in the light $V$ limit, which corresponds to the case C defined in subsection~\ref{sec:anatomy}. In this regime, our result is consistent with theirs.

Fig.~\ref{fig:SterileDM_Lmu-Ltau} also shows the experimental constraints, mostly adapted from Ref.~\cite{Bauer:2018onh}, where the colored shaded regions are already excluded and the regions enclosed by colored dashed curves will be covered by future experiments, correspondingly labeled. 
We also include the cosmological/astrophysical constraints from BBN and supernova 1987A~\cite{Escudero:2019gzq}, which mainly covers the region with $m_V$ below a few MeV scale.
Compared to the neutrinophilic case, the $U(1)_{L_\mu-L_\tau}$ model has a stronger interplay between the 
dark matter relic density favored regions
and experimental reaches. The present limits almost rule out the entire parameter space for the $m_4= 50\,{\rm keV}$ case. The reach of future experiments including SHiP~\cite{Anelli:2015pba,Alekhin:2015byh}, a dedicated NA62 analysis~\cite{Krnjaic:2019rsv}, and NA64$\mu$~\cite{Gninenko:2014pea,Gninenko:2018tlp}, if they occur, will be able to cover nearly the entire region above the $m_4 = 7.1\,{\rm keV}$ curve, that is, roughly the whole viable parameter space for dark matter relic density. 
The DUNE experiment, with a liquid argon near detector, can also probe the parameter space overlapping with the $(g-2)_\mu$ favored region, by measuring the neutrino trident production~\cite{Altmannshofer:2019zhy,Ballett:2019xoj}. 
DUNE, if equipped with a gaseous multi-purpose near detector, is also able to probe the displaced decay of $V$ into $e^+e^-$ via a loop-generated kinetic mixing~\cite{Berryman:2019dme}. 
Finally, a proposed muon-missing-momentum experiment~\cite{Berlin:2018bsc} can probe a parameter space similar to NA64$\mu$, for $m_V < 2m_\mu$.

\section{Gauged $U(1)_{B-L}$ Model}\label{sec:B-L}

The third model we consider is an extension of the Standard Model with $U(1)_{B-L}$ gauge symmetry~\cite{Davidson:1978pm,Mohapatra:1980qe}.
This is another popular framework where many phenomenological works have been done.
The interactions of the $B-L$ gauge boson $V$ take the form
\begin{equation}
g_{BL} V_\alpha \left(\frac{1}{3}\bar q \gamma^\alpha q - \bar \ell \gamma^\alpha \ell - \bar \nu \gamma^\alpha P_L \nu \right) \ ,
\end{equation}
universal for all quark and lepton flavors.
The new gauge interaction, under which active neutrinos are charged, could potentially facilitate the production of sterile neutrino dark matter. Like the $U(1)_{L_\mu-L_\tau}$ model, the active-sterile neutrino mixing has to be generated by a higher dimensional operator similar to Eq.~(\ref{eq:Sterile2}). 

The relic density calculation proceeds the same as before. The thermal potential is calculated using Eq.~(\ref{eq:VT}) with $\lambda_{\mu\mu}$ replaced by the gauge coupling $g_{BL}$. To calculate the thermal reaction rates of neutrinos, we make the same instantaneous decoupling approximation as described in the previous section.
The thermal reaction rate in the heavy mediator $V$ limit takes the form
\begin{equation}
\Gamma_V \approx \frac{7 \pi g_{BL}^4}{270 m_V^4} E \,T^4 \times
\left\{ 
\begin{array}{cl}
{\rm irrelevant}, & \hspace{1cm} T \gtrsim {\rm 1\,GeV} \\
22,&\hspace{1cm} 2m_\mu \leq T\lesssim {\rm 1\,GeV} \\
21,&\hspace{1cm} m_\mu \leq T< 2m_\mu \\
17,& \hspace{1cm} 2m_e\leq T<m_\mu \\
16,& \hspace{1cm} m_e\leq T<2 m_e \\
12,& \hspace{1cm} T<m_e\, . \\
\end{array}\right.
\end{equation}
For on-shell $V$ contribution, the rate is
\begin{equation}
\Gamma_{V} \simeq \frac{g_{BL}^4 m_V^2 T}{96\pi^2 \gamma_V E^2} \left( \ln \left( 1 + e^{\omega} \right) - \omega \rule{0mm}{4mm}\right) \,\times
\left\{ 
\begin{array}{cl} 
{\rm irrelevant},& \hspace{1cm} T\gtrsim {\rm 1\,GeV} \\
7,&\hspace{1cm} 2m_\mu \leq T\lesssim{\rm 1\,GeV} \\
5,& \hspace{1cm} 2m_e \leq T<2m_\mu \\
3,& \hspace{1cm} T<2m_e \\
\end{array}\right.
\end{equation}
where $w\equiv {m_V^2}/{(4ET)}$, and $\gamma_V$ is the decay width of the $B-L$ gauge boson~\cite{Heeck:2014zfa}.

\begin{figure}[t]
\centering\includegraphics[width=0.72\textwidth]{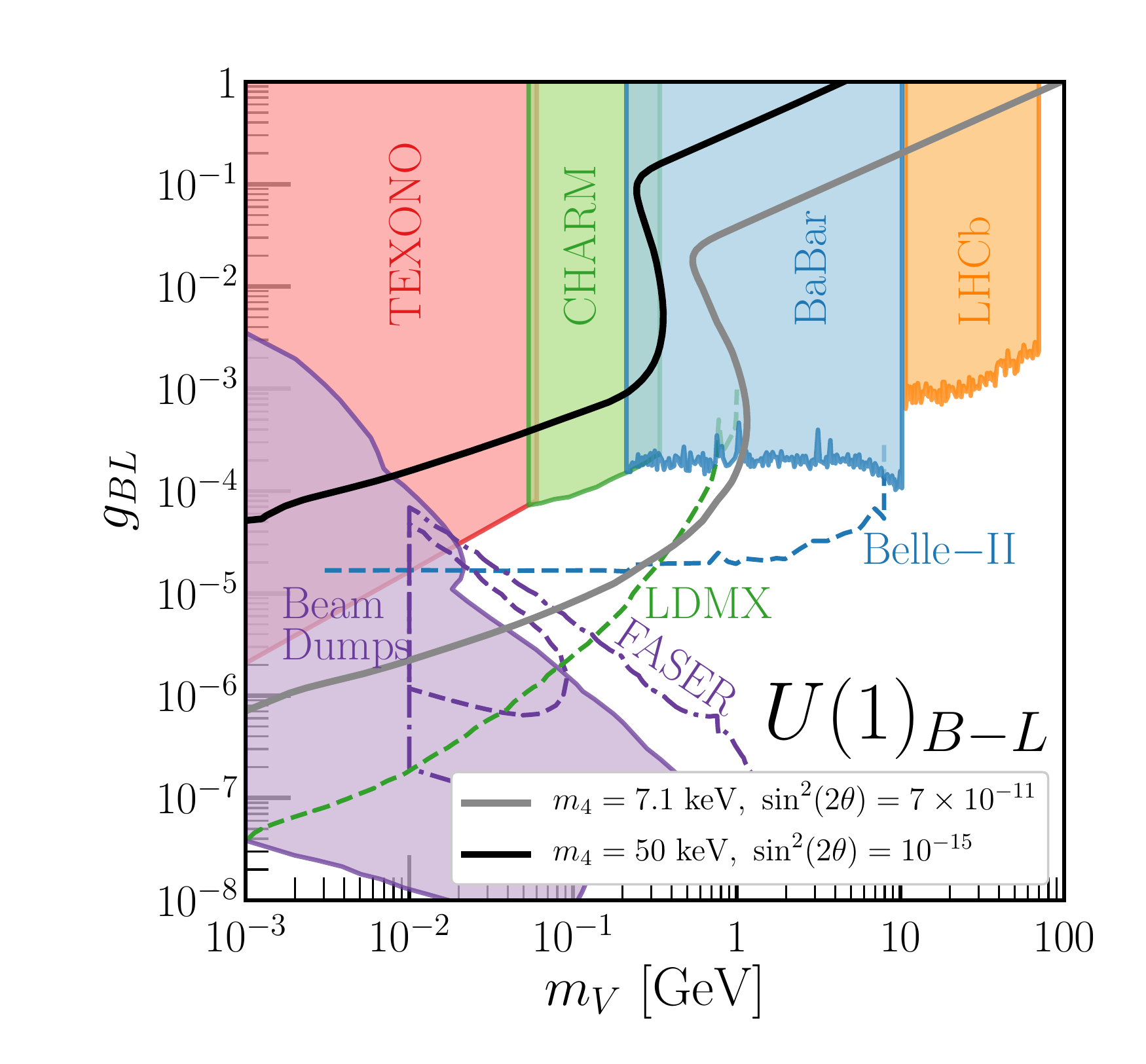}
\caption{Curves on the parameter space that yield the observed relic density of dark matter today for the $B-L$ vector boson model and $ m_4= 7.1\,{\rm keV},\,\sin^2(2\theta)=7\times 10^{-11}$ (gray), and  $m_4= 50\,{\rm keV},\,\sin^2(2\theta)=10^{-15}$ (black). Existing limits from a variety of probes are shown filled, colored regions: TEXONO (red), CHARM (green), BaBar (blue), LHCb (orange), and a set of beam dump experiments (purple). Future constraints from Belle-II (blue), LDMX (green), and FASER (purple) are shown as dashed lines.}
\label{fig:SterileDM_B-L}
\end{figure}

The above rates are calculated based on the interactions of neutrinos with themselves and the charged leptons.
For temperatures above GeV scale, the processes where neutrino scatters with/annihilates into baryons also contribute.\footnote{We neglect the neutrino-pion interactions which could occur through $V$-$\omega$-meson mixing.}
However, they are unimportant because the sterile neutrino dark matter is dominantly produced at temperatures well below a few hundred MeV. 
See discussion in subsection~\ref{sec:anatomy} and Fig.~\ref{fig:ABC}.
At higher temperatures, the effective active-sterile neutrino mixing angle is suppressed.

The result of our numerical calculation is presented in Fig.~\ref{fig:SterileDM_B-L}, where the gray and black curves corresponds to the two same choices of parameters as before:
$m_4= 7.1\,{\rm keV}$, $\sin^2(2\theta)=7\times 10^{-11}$ and
$m_4= 50\,{\rm keV}$, $\sin^2(2\theta)=10^{-15}$.
The relic density curves exhibit similar $\mathcal{S}$-shape, with three production regimes having distinctive parametric dependence, as discussed in subsection~\ref{sec:anatomy}.
In the same figure, we also show the experimental constraints on the $U(1)_{B-L}$ model parameter space~Ref.~\cite{Bauer:2018onh, Harnik:2012ni}, where the colored shaded regions are already excluded. 
Because the gauge boson $V$ couples to the electron, constraints are the strongest among the three models we have studied.
We find out that the entire curve in the $m_4= 50\,{\rm keV}$ case has already been firmly excluded.
The ongoing Belle-II~\cite{Kou:2018nap}, the upcoming FASER\footnote{The dashed purple line corresponds to FASER and the dot-dashed one to FASER 2. We use sensitivity projections from Ref.~\cite{Ariga:2018uku}, which correct an error present in Ref.~\cite{Bauer:2018onh}.} at LHC~\cite{Feng:2017uoz}, and the proposed LDMX experiment~\cite{Berlin:2018bsc} will cover the region enclosed by the blue and purple dashed curves. Combined, they will test all the relic density favored parameter space.
\footnote{Again, we remind that the $m_4 = 7.1\,$keV curve is roughly the lowest possible in this parameter space for which a sterile neutrino can satisfy the observed relic abundance of dark matter.}

To summarize, the available parameter space for sterile neutrino dark matter production is already quite small in the $U(1)_{B-L}$ model. It can be fully covered by experiments in the near future.

\section{Conclusion}\label{sec:conclude}

In this work we considered a sterile neutrino as the dark matter candidate, and studied the origin of its relic abundance. 
The Dodelson-Widrow mechanism fails to work, given the severe constraints from X-ray indirect detection and small scale structure observations. 
We explored simple extensions of the Standard Model where a new vector boson $V$ mediates new interactions for active neutrinos that facilitates efficient production of sterile neutrino dark matter. We calculated the dark matter production via neutrino scattering and oscillation in the early universe and identify three scenarios (A, B, C as defined section~\ref{sec:anatomy}). 
Depending whether $V$ plays the role of a heavy or light mediator, it leads to different parametric dependence in the final dark matter relic density. 
The viable mass of $V$ lies between MeV to GeV and the corresponding neutrino coupling ranges from $10^{-6}$ to $10^{-2}$.
For comparison, we also carried out similar calculations in the gauged $U(1)_{L_\mu-L_\tau}$ and $U(1)_{B-L}$ models. The main results are presented in Figs.~\ref{fig:SterileDM_NuPhil}, \ref{fig:SterileDM_Lmu-Ltau} and \ref{fig:SterileDM_B-L}.

The parameter space for viable dark matter relic density found in this work is a well motivated target for the next experimental probes.
Among the three models studied, the $U(1)_{B-L}$ model is the most strongly constrained. Its parameter space for sterile neutrino dark matter is already narrow and will be fully covered by future high intensity experiments, including BELLE-II, FASER, and LDMX. The $U(1)_{L_\mu-L_\tau}$ case is relatively less constrained but could also be covered by the proposed experiments, including SHiP, NA62, NA64-$\mu$, $M^3$, and DUNE.
In contrast, the model with a neutrinophilic vector boson is the least constrained.  The strongest present limit comes from the Higgs boson invisible decay search, which will be improved by HL-LHC. A portion of its parameter space has not been targeted by any experiment, to our knowledge.

On the theory side, the same dark matter production mechanisms could also work in, e.g., gauged $U(1)_{L_{e}-L_\mu}$ or $U(1)_{L_{e}-L_\tau}$ models, or even gauged lepton-number that has other motivations~\cite{Carena:2018cjh, Carena:2019xrr}. The presence of electron coupling for the vector boson $V$ in these models implies strong experimental probes similar to $U(1)_{B-L}$.

\bigskip
\section*{Acknowledgments}
We would like to thank Andr\'e de Gouv\^ea for helpful discussions and Felix Kling for useful comments regarding FASER.
KJK is supported by Fermi Research Alliance, LLC under Contract No. DE-AC02-07CH11359 with the U.S. Department of Energy, Office of Science, Office of High Energy Physics.
MS acknowledges support from the National Science Foundation, Grant PHY-1630782, and to the Heising-Simons Foundation, Grant 2017-228. The research of WT is supported by the College of Arts and Sciences of Loyola University Chicago. The work of YZ is supported by the Arthur B. McDonald Canadian Astroparticle Physics Research Institute.

\bibliography{DWZprime}
\end{document}